\documentclass[12pt,preprint]{aastex}
\usepackage{amssymb}
\usepackage{amsmath}
\usepackage{epsfig}

\newcommand{\myemail}{racusin@astro.psu.edu}
\newcommand{\swift}{{\it Swift }}
\newcommand{\swiftnsp}{{\it Swift}}

\shorttitle{Jet breaks and Energetics of {\it Swift} GRB X-ray Afterglows}
\shortauthors{Racusin et al.}

\begin{document}

\title{Jet breaks and Energetics of {\it Swift} GRB X-ray Afterglows}

\author{J. L. Racusin\altaffilmark{1}, E. W. Liang\altaffilmark{2,3},
  D. N. Burrows\altaffilmark{1},
  A. Falcone\altaffilmark{1}, T. Sakamoto\altaffilmark{4},
  B. B. Zhang\altaffilmark{2}, B. Zhang\altaffilmark{2}, 
  P. Evans\altaffilmark{5}, J. Osborne\altaffilmark{5}}
\altaffiltext{1}{Department of Astronomy \& Astrophysics, The Pennsylvania
  State University, 525 Davey Lab, University Park, PA 16802, USA; \myemail}
\altaffiltext{2}{Department of Physics, University of Nevada, Las Vegas, NV
  89154, USA}
\altaffiltext{3}{Department of Physics, Guangxi University, Nanning 530004,
  China}
\altaffiltext{4}{Astrophysics Science Division, Code 661, NASA's Goddard
  Space Flight Center, 8800 Greenbelt Road, Greenbelt Maryland 20771, USA}
\altaffiltext{5}{Department of Physics and Astronomy, University of Leicester,
  Leicester, LE1 7RH, UK}

\begin{abstract}
  We present a systematic temporal and spectral study of all \swift-XRT
  observations of GRB afterglows discovered between 2005 January and 2007
  December.  After constructing and fitting all light curves and spectra to
  power-law models, we classify the components of each afterglow in terms
  of the canonical X-ray afterglow and test them against the
  closure relations of the forward shock models for a variety of parameter
  combinations.  The 
  closure relations are used to identify potential jet breaks with
  characteristics including the 
  uniform jet model with and without lateral spreading and energy injection, and
  a power-law structured jet model, all with a range of parameters.
  With this technique, we survey the X-ray afterglows with strong evidence for
  jet breaks ($\sim 12\%$ of our sample), and reveal cases of potential jet
  breaks that do not appear plainly from the light curve alone (another
  $\sim30\%$), leading to insight into the missing jet break
  problem.  Those X-ray light curves that do not show breaks or have breaks that
  are not consistent with one of the jet models are explored to place
  limits on the times of unseen jet breaks.  The distribution of jet
  break times ranges from a few hours to a few weeks with a median of
  $\sim 1$ day, similar to what was found pre-\swiftnsp.  
  On average \swift GRBs have lower isotropic 
  equivalent $\gamma$-ray energies, which in turn results in lower
  collimation corrected $\gamma$-ray energies than those of pre-\swift
  GRBs.  Finally, we explore the implications for GRB jet geometry and
  energetics. 

\end{abstract}


\section{Introduction}


One of the most surprising puzzles to emerge from the \swift \markcite{gehrels04}({Gehrels} {et~al.} 2004)
mission and its dynamic study of Gamma-ray Bursts (GRBs) is the lack of
expected jet breaks in X-ray afterglow emission.  It is vital to the
entire study of GRBs to understand jet geometry, because of the inferred effects on
the total output energy, GRB rate, afterglow structure, interactions with
environment, and jet physical mechanisms. Pre-\swift optical observations
showed tens of cases of steepening in the light curves several days
after the GRB triggers \markcite{frail01,bloom01,zeh06}({Frail} {et~al.} 2001; {Bloom}, {Frail}, \& {Sari} 2001; {Zeh}, {Klose}, \& {Kann} 2006).  This steepening was
interpreted as evidence for the collimation of the burst ejecta 
with physical half-angle $\theta_j$.  The ejecta moves
at relativistic velocities with a bulk Lorentz factor, $\Gamma$,
and the radiation is relativistically beamed into an angle
$\theta=1/\Gamma$.  
As the ejecta sweeps up surrounding material, the fireball
decelerates, with the beaming angle eventually exceeding the physical
collimation angle, causing a sudden increase in the rate of decay of
the flux (i.e. the jet break).  At the same time, sideways expansion of the
ejecta with relativistic speeds also causes a sudden flux decrease
\markcite{sari99,rhoads99,zhang04}({Sari}, {Piran}, \& {Halpern} 1999; {Rhoads} 1999; {Zhang} \& {M{\'e}sz{\'a}ros} 2004). Most likely both of these effects
contribute to the jet breaks, and therefore, both models must be considered.
Breaks are expected to be achromatic based on the assumption that the
afterglow emission regions and mechanisms are the same for various spectral
regimes, and should therefore only reflect ejecta geometry.
Achromaticity has indeed been confirmed in the optical/near-infrared
bands in pre-\swift GRBs \markcite{kulkarni99,harrison01,klose04}({Kulkarni} {et~al.} 1999; {Harrison} {et~al.} 2001; {Klose} {et~al.} 2004). 

\begin{figure}
  \begin{center}
  \includegraphics[scale=0.5,angle=90]{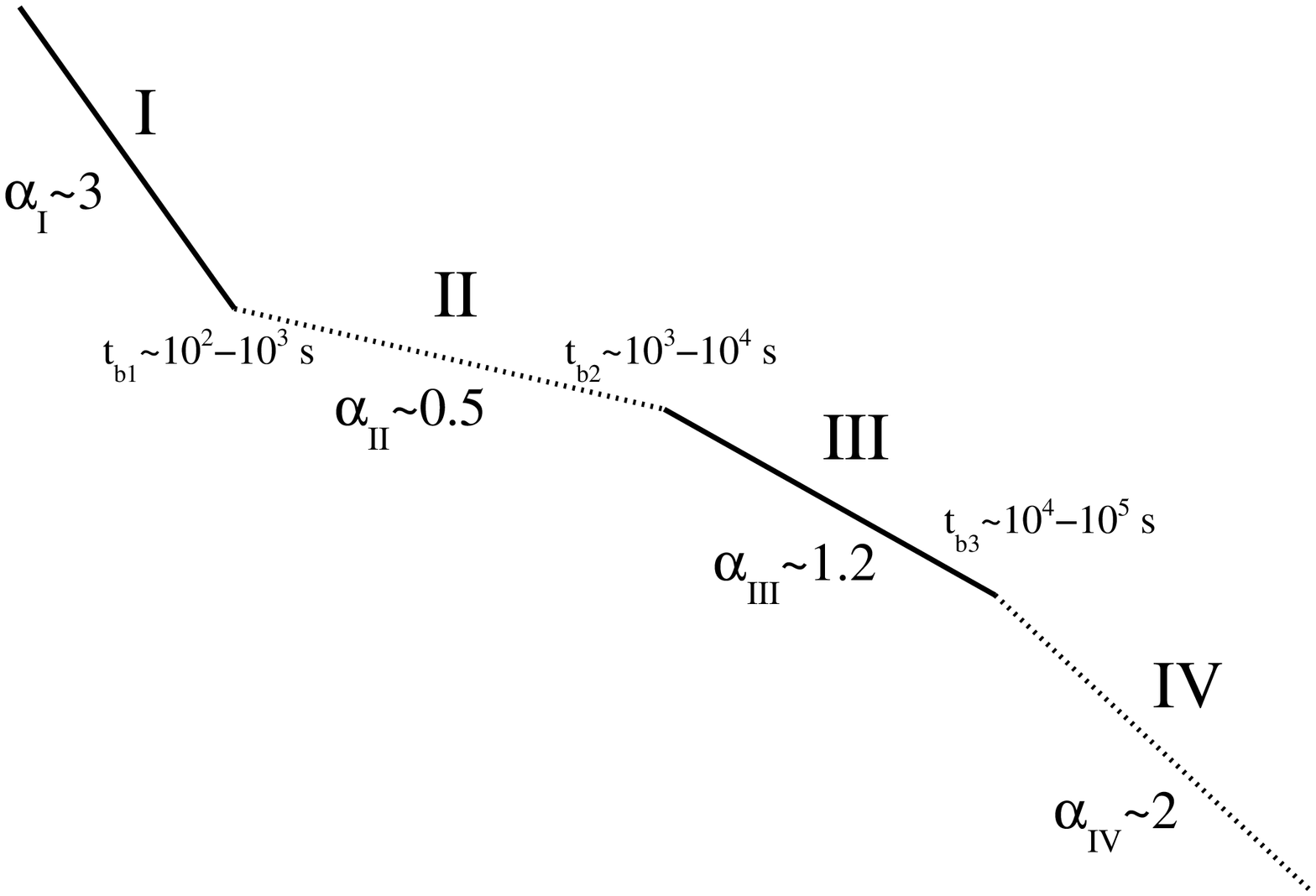}
  \end{center}
  \caption{Canonical X-ray afterglow light curve defined by
    \markcite{zhang06}{Zhang} {et~al.} (2006) and \markcite{nousek06}{Nousek} {et~al.} (2006).  Segment I is generally attributed to
    high latitude emission.  Segment II is due to continuous energy
    injection by the central engine.  Segment III is the normal
    spherical decay of the afterglow.  Segment IV is the post-jet
    break decay.  Segment V (not shown) is due to flares which can occur
    during any phase, in multiple, and in widely varying strengths.
 \label{fig:canon}}
\end{figure}

In the \swift era, it is X-ray afterglow light curves that provide the most
homogeneous data set to study GRB afterglows.  With the rapid GRB
triggers provided by the \swift-BAT, and the autonomous prompt \swift-XRT
observations that frequently begin within 1-2 minutes of the trigger, X-ray
afterglows have gone from sparsely sampled single power-laws (Beppo-SAX era,
\markcite{depasquale06}{de Pasquale} {et~al.} 2006) to a rich database of light curves with widely varying
properties and durations.  A common canonical shape of the \swift-XRT X-ray
light curves emerged (\markcite{nousek06,zhang06}{Nousek} {et~al.} (2006); {Zhang} {et~al.} (2006), Figure \ref{fig:canon})
with five components (an 
initial steep decay, a shallow-decay ``plateau'' phase, a normal decay, a
jet-like decay component, and flares) that could be used to explain the
overall structure of 
the afterglows.  While elements of this canonical
picture are seen in most X-ray afterglows, few afterglows contain all 5
components. Surprisingly, fully 
``canonical'' jet breaks are rarely observed in the XRT light curves
\markcite{burrows07,liang08,evans09}({Burrows} \& {Racusin} 2007; {Liang} {et~al.} 2008; {Evans} {et~al.} 2008).

In this work, we assume that all X-ray afterglows have inherently similar
shapes (the canonical shape) with deviations in behavior due to environment,
electron spectral shape, spectral 
regime, presence of energy injection, and jet properties.  We also explore
observational biases such as late beginning and early ending of 
observation, flares, and observing gaps that lead to missing portions of 
individual light curves which can produce ambiguities in the
identification of segments.

We assume that the light curve segments following the initial steep decay (Figure
\ref{fig:canon}, segments II-IV) are due to external forward shocks.
Although we recognise that several alternative models have recently been
proposed to explain the origin of the X-ray afterglows 
(e.g. \markcite{genet07,shao07,ghisellini07,liang07,uhm07,liang08,panaitescu08,kumar08}{Genet}, {Daigne}, \& {Mochkovitch} (2007); {Shao} \& {Dai} (2007); {Ghisellini} {et~al.} (2007a); {Liang}, {Zhang}, \& {Zhang} (2007); {Uhm} \& {Beloborodov} (2007); {Liang} {et~al.} (2008); {Panaitescu} (2008); {Kumar}, {Narayan}, \& {Johnson} (2008)),
as discussed below (see also \markcite{liang07,liang08}{Liang} {et~al.} 2007, 2008), the X-ray data can be
generally interpreted within the framework of the forward shock, so that
invoking non-forward-shock models is not absolutely demanded by the data.
We also focus solely on the X-ray behavior of the afterglow light curves,
while acknowledging that chromatic behavior inferred from optical
observations provides important clues into jet properties
\markcite{liang08}({Liang} {et~al.} 2008).

Based on the optical afterglow observations in the pre-\swift era 
\markcite{frail01,bloom03}({Frail} {et~al.} 2001; {Bloom}, {Frail}, \& {Kulkarni} 2003), we expected to find jet breaks occurring within
several days after the bursts, with the light curves breaking to decay slopes
of $\sim 2.2$ \markcite{sari99}({Sari} {et~al.} 1999). Several recent studies 
\markcite{burrows07,liang08,kocevski07,willingale07,evans09}({Burrows} \& {Racusin} 2007; {Liang} {et~al.} 2008; {Kocevski} \& {Butler} 2008; {Willingale} {et~al.} 2007; {Evans} {et~al.} 2008) have searched for jet
breaks 
in the XRT data and agree that there is a substantial deficit
relative to pre-{\it Swift} expectations.  \markcite{panaitescu07}{Panaitescu} (2007) suggests 
additional potential jet breaks in the sample, but is very broad in his jet
break definition, attributing even breaks with shallower decays occurring
after plateaus to jet breaks without discussion of the more
global context of the light curves.  This 
method does not follow the framework of the canonical picture observed in many
afterglows, though it suggests that some jet breaks are buried in the existing
measured breaks, which we explore in more detail.  Previous studies of large 
afterglow samples have applied only the simplest afterglow models, rather than
the detailed interpretations needed to explain GRBs in 
individual cases.  For example, the end of the plateau phase is often
attributed to a cessation of energy injection without considering the
possibility that it might represent a jet break during energy injection (see
also \markcite{depasquale08}{de Pasquale} {et~al.} 2008). Other model variations including a flat
electron spectrum, different progenitor environments, and jet geometry and
dynamics would slightly alter the properties of the canonical behavior.
Therefore, in this study we perform a more generalized characterization of all
XRT afterglows, reexamining a wide variety of closure relations to evaluate
which segments of each light curve are consistent with each family of
closure relations, whether there are any jet breaks that have been
previously misinterpreted, and what limits can be placed on jet break
statistics and energetics.

There are multiple reasons that we may not be detecting jet breaks in XRT
data. We ask the following questions to explore this problem:  Are jet breaks
subtle and buried within our observing errors (see also
\markcite{curran08}{Curran}, {van der Horst}, \&  {Wijers} 2008)?  Do the jet breaks occur after 
the XRT observations end?  Do observational biases that cause us to miss
parts of the light curves result in ambiguous classifications?  Could some of
the breaks at the end of the plateau phases actually be jet breaks
that are masked by continuing energy injection?  Are those GRBs for which jet 
breaks are not detected somehow intrinsically or observationally different
than those for which they are detected?  The goal of this 
study is to attempt to answer these questions. 

We describe the data sample selection, temporal analysis, and spectral analysis
in \S\ref{sec:data}, the closure relations in
\S\ref{sec:cr}, the results in \S\ref{sec:results}, discussion and implications
in \S\ref{sec:disc}, and conclude in \S\ref{sec:conc}. Throughout this paper,
we adopted the convention $F \propto t^{-\alpha}\nu^{-\beta}$ where $\alpha$ is
the temporal index and $\beta$ is the spectral index, $F$ is the energy flux
(with cgs units of $erg\ cm^{-2}\ s^{-1}$), and we use cosmological
parameters $H_0 = 70~{\mathrm{km~s^{-1}~Mpc^{-1}}}$, $\Omega_M=0.3$,
$\Omega_{\Lambda}=0.7$.

\section{Data Reduction} \label{sec:data}

The \swift-XRT observed 262 GRB X-ray afterglows between
2005 January and 2007 December.  Our sample also contains 21 bursts discovered
by Integral, HETE, or the Interplanetary Network (IPN) that were
followed up by the \swift-XRT beginning 
within approximately one day. We include only those afterglows with at least
60 background-subtracted photons, enough to construct a basic light curve and
characterize temporal and spectral properties (in the methods described in
Sections \S\ref{sec:temp} and \S\ref{sec:spec}).  
Removing those objects for which we do not have adequate temporal and
spectral information, our resulting main sample consists of 230 GRB X-ray
afterglows; 15 of those afterglows were not originally discovered by
{\it Swift}; 13 are short bursts ($T_{90}<2$ sec); and 85 have measured redshifts
reported in the literature as indicated 
in Tables \ref{table:prominent}-\ref{table:unlikely}.

Level 1 data products were downloaded from the NASA/GSFC \swift Data Center
(SDC) and processed using XRTDAS software (v2.0.1).  The {\it xrtpipeline} task
was used to generate level 2 cleaned event files.  Only events with Windowed
Timing (WT) mode grades 0-2 and Photon Counting (PC) mode grades
0-12 and energies between $0.3-10.0$ keV were used in subsequent temporal and
spectral analysis.

\subsection{Temporal Analysis} \label{sec:temp}

We assume that all X-ray light curves in our sample inherently follow the
canonical form (Figure \ref{fig:canon}) described by \markcite{zhang06}{Zhang} {et~al.} (2006) and
\markcite{nousek06}{Nousek} {et~al.} (2006).  
These 4 segments and additional component are I: the
initial steep decay often attributed to high-latitude emission or the
curvature effect \markcite{kumar00,qin04,liang06,zhang07a}({Kumar} \& {Panaitescu} 2000; {Qin} {et~al.} 2004; {Liang} {et~al.} 2006; {Zhang}, {Liang}, \&  {Zhang} 2007b); II: the plateau, 
which is frequently attributed to continuous energy injection from
the central engine
\markcite{rees98,dai98,sari00,zhang01,granot06,zhang06,liang07}({Rees} \& {M{\'e}sz{\'a}ros} 1998; {Dai} \& {Lu} 1998; {Sari} \& {M{\'e}sz{\'a}ros} 2000; {Zhang} \& {M{\'e}sz{\'a}ros} 2001; {Granot} \& {Kumar} 2006; {Zhang} {et~al.} 2006; {Liang} {et~al.} 2007);
III: the normal decay due to the deceleration of an adiabatic fireball
\markcite{meszaros02,zhang06}({M{\'e}sz{\'a}ros} 2002; {Zhang} {et~al.} 2006); IV: the post-jet break phase 
\markcite{rhoads99,sari99,meszaros02,piran05}({Rhoads} 1999; {Sari} {et~al.} 1999; {M{\'e}sz{\'a}ros} 2002; {Piran} 2005); V: flares, which are 
seen in $\sim 1/3$ of all \swift GRB X-ray afterglows during any phase (I-IV)
and are believed to be caused by sporadic emission from the
central engine \markcite{burrows05,zhang06,chincarini07,falcone07}({Burrows} {et~al.} 2005; {Zhang} {et~al.} 2006; {Chincarini} {et~al.} 2007; {Falcone} {et~al.} 2007).  We
classify the data into these segments based upon the criteria described below. 
Only 25 cases contain all 4 light curve segments, with 14 cases also
containing flares.

This analysis does not address the phenomenon of X-ray flares, but rather
excludes them from the spectral and temporal analysis.  See
\markcite{chincarini07}{Chincarini} {et~al.} (2007) and \markcite{falcone07}{Falcone} {et~al.} (2007) for detailed studies of X-ray
flares and analysis on this data set.
These studies have shown that significant spectral evolution occurs
throughout the flares, therefore in order to constrain the properties of the
underlying afterglows, we remove the time intervals of significant flaring
from subsequent temporal and spectral analysis.  Flaring was determined by
visual inspection of the light curves and hardness ratios.  Only the most
apparent flares were removed, with no attempt to constrain small scale
or micro-flaring.  Large flares that overlap and significantly exceed the
level of the underlying afterglow can also mask whole segments, making it
impossible to determine the underlying temporal and spectral properties.
The flaring in these cases usually occurs at the beginning of light curves,
overwhelming segment I and leading to light curves with apparent segments II-III 
afterwards.  Rather than guessing the properties or presence of these
specific flaring
segments, we remove those time intervals and proceed as if they were not
part of the rest of the light curves.

For those X-ray afterglows that were also observed by {\it Chandra} at late
times, we include those data points in our temporal but not spectral fits.
These bursts include GRB 051221A \markcite{burrows06}({Burrows} {et~al.} 2006), GRB 050724 \markcite{grupe06}({Grupe} {et~al.} 2006),
and GRB 060729 (Grupe et al., in-preparation).

\subsubsection{Light Curve Construction} \label{sec:lccon}
All light curves were extracted from the public XRT-team light curve repository
\markcite{evans07,evans09}({Evans} {et~al.} 2007, 2008).  These XRT light curves were created by extracting the
counts in a circular region around the afterglow with a variable source
extraction radius designed to optimize the S/N ratio depending on the count rate in
both the Windowed Timing (WT) and Photon Counting (PC) mode data.  A region
clear of any serendipitous background sources was used to
estimate the contribution of background counts in the source region.
The number of counts per bin are chosen depending on the count rate to show
sufficient detail with reasonable error bars and binning.  The
background-subtracted count rates are also corrected for the portion of the
PSF excluded 
by the extraction region and any proximity to bad columns and hot pixels in the
XRT CCD. Where necessary, corrections for photon pile-up were also made by
excluding a central portion of the extraction region.

\subsubsection{Light Curve Fitting} \label{sec:lcfit}

We have developed tools to fit single power-laws, broken power-laws, double broken
power-laws, and triple broken power-laws with the following functional forms to
the XRT light curves.
\\
\\
single power-law:
\begin{equation}\label{eq:spl}
  F(t)=N\,t^{-\alpha_0}
\end{equation}
broken power-law:
\begin{equation}\label{eq:bpl}
  F(t)=N
  \begin{cases}
    t^{-\alpha_1} & t<t_{b_1} \\
    t_{b_1}^{(\alpha_2-\alpha_1)}\,t^{-\alpha_2} & t>t_{b_1} \\
  \end{cases}
\end{equation}
double broken power-law:
\begin{equation}\label{eq:b2pl}
  F(t)=N
  \begin{cases}
    t^{-\alpha_1} & t<t_{b_1} \\
    t_{b_1}^{(\alpha_2-\alpha_1)}\,t^{-\alpha_2} & t_{b_1}<t<t_{b_2} \\
    t_{b_1}^{(\alpha_2-\alpha_1)}\,t_{b_2}^{(\alpha_3-\alpha_2)}\,t^{-\alpha_3} &
    t>t_{b_2} \\
  \end{cases}
\end{equation}
triple broken power-law:
\begin{equation}\label{eq:b3pl}
  F(t)=N
  \begin{cases}
    t^{-\alpha_1} & t<t_{b_1} \\
    t_{b_1}^{(\alpha_2-\alpha_1)}\,t^{-\alpha_2} & t_{b_1}<t<t_{b_2} \\
    t_{b_1}^{(\alpha_2-\alpha_1)}\,t_{b_2}^{(\alpha_3-\alpha_2)}\,t^{-\alpha_3} &
    t_{b_2}<t<t_{b_3} \\
    t_{b_1}^{(\alpha_2-\alpha_1)}\,t_{b_2}^{(\alpha_3-\alpha_2)}\,t_{b_3}^{(\alpha_4-\alpha_3)}\,t^{-\alpha_4}
& t>t_{b_3} \\
  \end{cases}
\end{equation}
where $N$ is the normalization, $t$ is the time since the burst
trigger, $F(t)$ is the count rate over the soft X-ray band ($0.3-10$ keV),
$t_{b_1,b_2,b_3}$ are the times of breaks in the  
light curves, and $\alpha_{(0,1,2,3,4)}$ are the temporal indices of the power-law
fits.

Our software, written in IDL, requires user input for initial guesses of the
location and number of breaks (and therefore power-law model).  Based upon
visual inspection, the user first eliminates all obvious time intervals with
significant flaring.\footnote{We note that broad-peaked flares, poorly sampled flares,
or the sum of many flares could be misinterpreted as a single power-law
decay from one of the phases.} The user then makes initial guesses for
light curve break 
times, and a least-squares fitting routine is used to fit each model.  When 
the addition or removal of light curve segments from the initial model
also provides an adequate fit, we perform an F-test and if the fit is 
improved at a $99\%$ confidence level, then the new model is retained.

In order to accurately measure the light curve model parameter errors
without over-estimating them, we tested both $\Delta 
\chi^2$ confidence interval mapping and Monte Carlo simulation methods.  
In the Monte Carlo method we created 10000 simulated light curves for each
GRB light curve jiggling the data points by an amount drawn randomly from
the Poisson distributions derived from the source and background counts.
Each of these light curves was fit with the same method as the real light
curves. The $90\%$ and $2\sigma$ confidence intervals were taken from the
distributions of each fit parameter from the simulations.  The broken power-law fits
were not well behaved in $\Delta \chi^2$ contour space due to 
data binning, light curve gaps, and logarithmic fits, resulting in larger
error estimates compared to the Monte Carlo method.  The latter method is
more free from assumptions and biases, therefore we chose to use the
Monte Carlo light curve parameter error estimates for the following analysis.

We classify each segment of the light curves in terms of the canonical form
(Figure \ref{fig:canon}) using the the following criteria.
If the light curve is a triple broken power-law then identification of its
segments is unambiguous, and it is designated as a type
I-II-III-IV.  If the light curve is best fit by a double broken power-law, we
apply the criteria that if 
$\alpha_1>\alpha_2$ then it is designated as segments I-II-III or if
$\alpha_1<\alpha_2$ then it is designated as segments II-III-IV.
If the light curve is best fit by a singly broken power-law, we apply the
criteria that 
if $\alpha_1>\alpha_2$ then it is designated as segments I-II, while if
$\alpha_1<\alpha_2$ then it can be interpreted as either segments II-III or
III-IV.  If the light curve is best fit by a single power-law then any segment
(I,II,III,IV) is possible. 

\begin{figure}
  \begin{center}
    \includegraphics[scale=0.45]{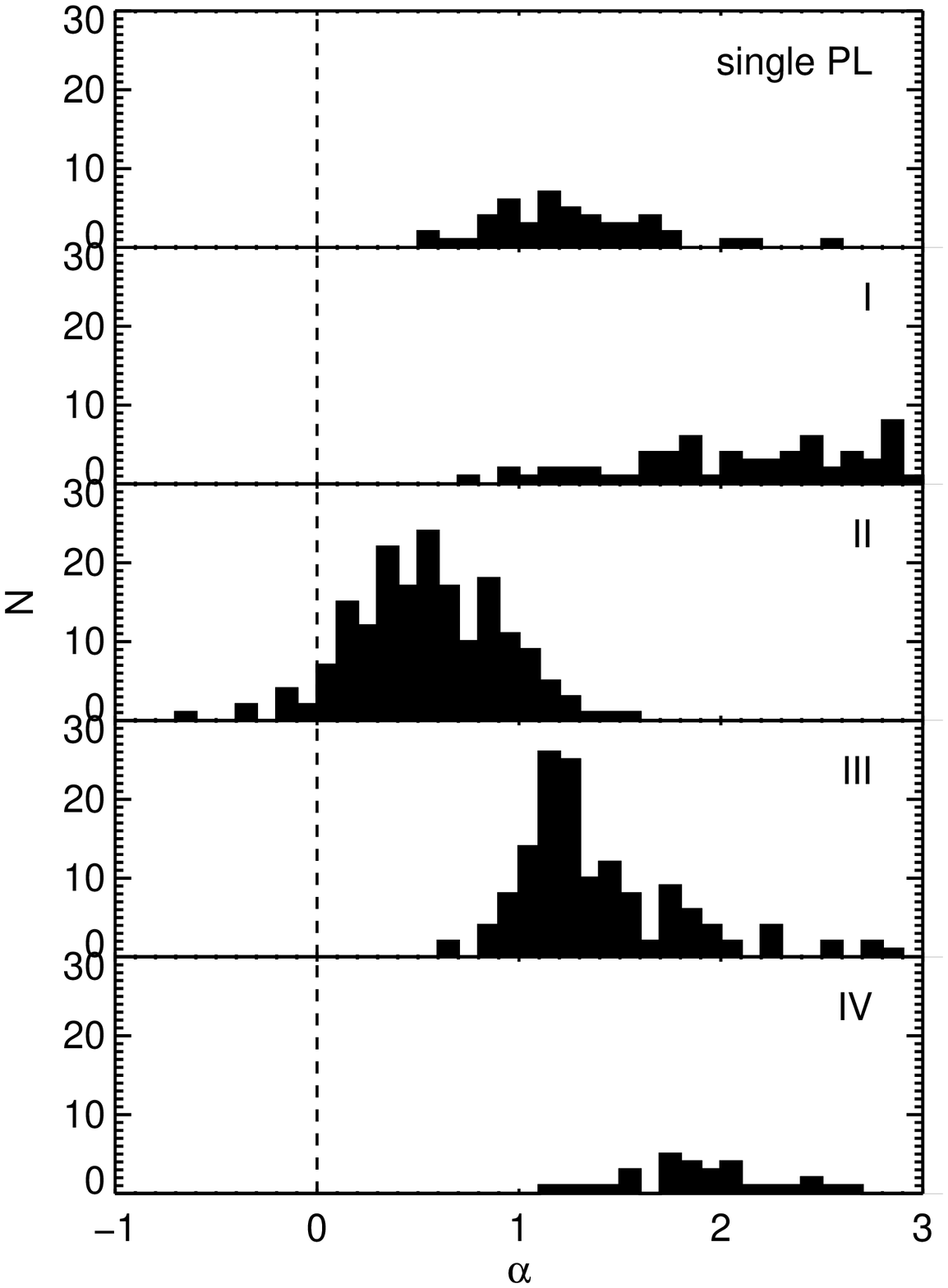}
    \includegraphics[scale=0.45]{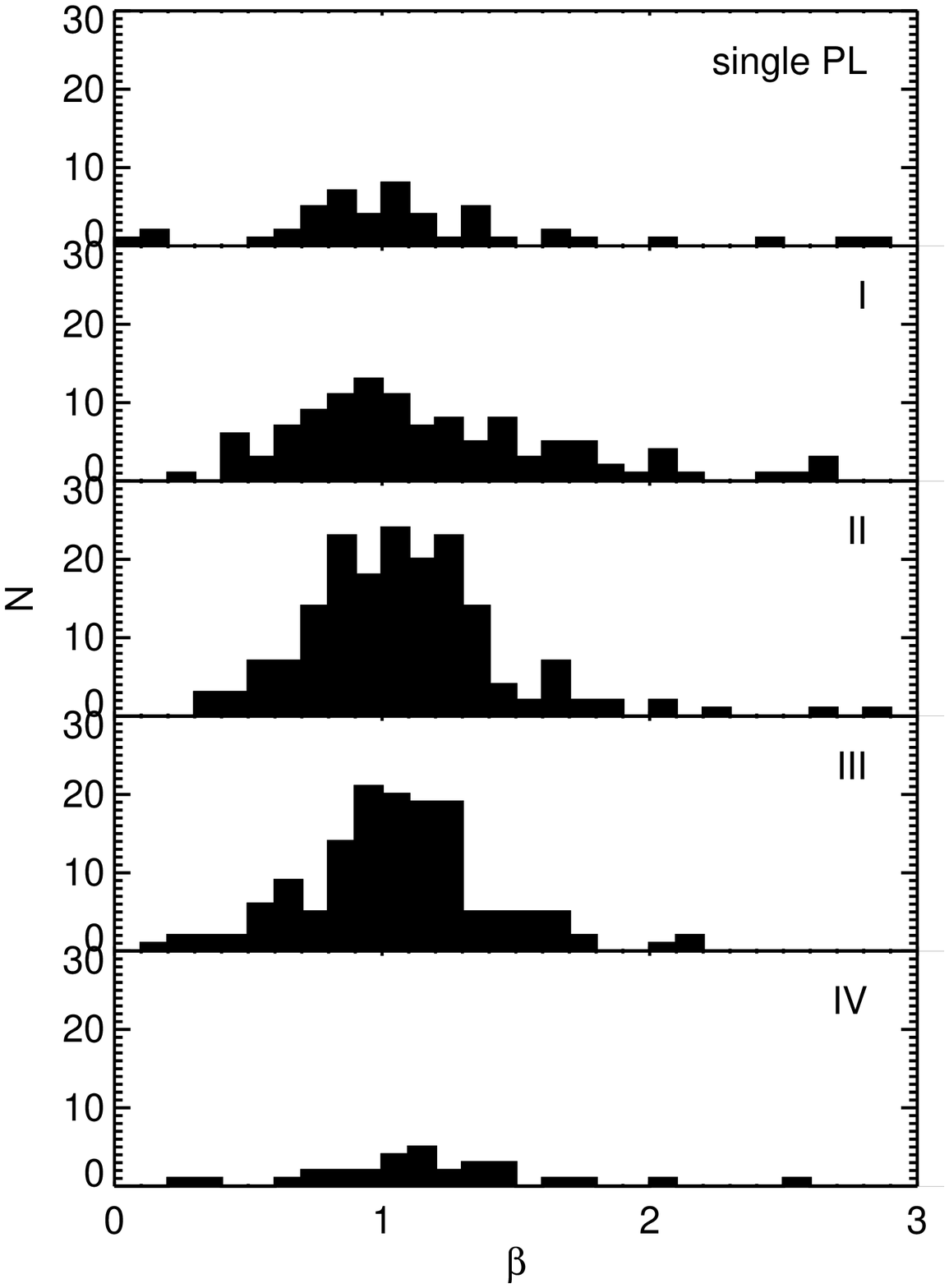}
  \end{center}
  \caption{Distributions of light curve decay indices ($\alpha$, {\it left}).
    and corresponding spectral indices ($\beta$, {\it right}). The top
    panel in each figure shows the single power-law cases, 
    while the other panels are split into light curve segments as identified in
    Figure  \ref{fig:canon}.  Note that
    the overlap in the temporal distributions of segments II and III
    are due to contamination from ambiguous light curves as described
    in \S\ref{sec:feasible}. The dashed line at $\alpha=0$ indicates
    the distinction between rising and decaying light
    curves. \label{fig:abfits}}  
\end{figure}

The distributions of these temporal fits are given in the left panel of
Figure \ref{fig:abfits}. 
Although the $\alpha$ distributions are broad, the
different segments are clearly separated by our classification criteria
described above.  Kolmogorov-Smirnov (KS) tests show that the segments I through IV
are different at $>99.9\%$ level.  The single power-law distribution also
differs from segments I, II, and IV at $>99.9\%$ level.  However, the single
power-law and segment III 
distributions are more similar ($4\%$ probability of begin drawn from same
inherent distribution), suggesting that some of the single
power-laws are segments III, with others drawn from segments I, II, and IV.

This scheme has no implicit constraints on the range of temporal indices, but
does leave some ambiguity in the case of a broken power-law with
$\alpha_1<\alpha_2$.  This case is equivocal between segments II-III and
III-IV.  We generally assume that they are cases of II-III when looking at
sample distributions.  However, we still fit the post-jet break closure
relations, 
allowing for the possibility of III-IV.  This ambiguity and distinguishing
criteria are further addressed in \S\ref{sec:feasible}.


\subsection{Spectral Analysis} \label{sec:spec}

We have constructed spectra for each segment of each light curve
distinguished using the temporal fits defined in \S\ref{sec:lcfit}.  These
spectra were extracted using XSELECT with 20 pixel radius source extraction
region, and a 40 pixel radius background region. All analysis used
the version 011 (release date 2008-05-14) response matrices from the \swift
CALDB, and Ancillary Response Files (ARFs) were made using
the {\it xrtmkarf} task.  The spectra were grouped using the {\it grppha}
task with 20 counts per bin and were fit using $\chi^2$ statistics 
unless there were fewer than 150 counts, in which case they were grouped with
10 counts per bin and were fit using Cash statistics.
We fit these spectra in
XSPEC to an absorbed power-law with two absorption components, one fixed to
the Galactic value \markcite{kalberla05}({Kalberla} {et~al.} 2005), and another freely varying using the
measured redshift if available. 

The photon indices ($\Gamma$) of the spectral fits are used to measure $\beta$,
the energy spectral index, where $\beta=\Gamma-1$.  The distributions of
these spectral fits are shown in  Figure \ref{fig:abfits}.  Segments II through
IV are statistically similar (as tested with a KS test).  Of
those light curves with a segment IV, $\sim 90\%$ are
consistent with minimal or no spectral evolution through segments II, III, and
IV ($90\%$ confidence errors). The KS test shows that segment I differs from
segments II and III at $>98\%$ level, with the distribution peaking at a
slightly lower $\beta$, consistent with the possibility of a different
physical origin of this phase \markcite{zhang06,liang06,zhang07a,zhang09}({Zhang} {et~al.} 2006; {Liang} {et~al.} 2006; {Zhang} {et~al.} 2007b, 2009).  
The distribution of $\beta$s for the single power-law light curves is
statistically consistent with the other individual segments.  These spectral
properties are in agreement with the suggestion from the temporal distributions
that this sample is a mixture of the other segments.

\section{Closure Relations} \label{sec:cr}
The fireball model predicts the evolution 
of the spectral and temporal properties of GRB afterglows as the external shocks
decelerate in the external environment.  These
effects can be characterized by relationships between the temporal and spectral
indices ($\alpha$ and $\beta$).  These so-called ``closure relations''
of the fireball model depend on the physical processes in the relevant portion of the
afterglow light curve, the surrounding environment, electron spectral index,
cooling regime, and jet geometry
(\markcite{rees94,meszaros97,sari98,chevalier00,dai01}{Rees} \& {M{\'e}sz{\'a}ros}, 1994; {M{\'e}sz{\'a}ros} \& {Rees}, 1997; {Sari}, {Piran}, \& {Narayan}, 1998; {Chevalier} \& {Li}, 2000; {Dai} \& {Cheng}, 2001, 
see reviews by \markcite{meszaros02,zhang04,piran05}{M{\'e}sz{\'a}ros}, 2002; {Zhang} \& {M{\'e}sz{\'a}ros}, 2004; {Piran}, 2005). 
We apply a large set of possible 
models to each segment of the light curves and use them in
conjunction to narrow the list of possible physical models to explain each
afterglow segment, assuming that synchrotron radiation is the dominant mechanism and
including only the corresponding relevant relations.  We also use them to
determine the presence and properties of jet breaks in our sample of X-ray light
curves.  All of the analytical closure relations assume a simple spectrum
with sharp breaks, whereas in reality these breaks are likely smooth
\markcite{granot02}({Granot} \& {Sari} 2002).  However, unfortunately even with broadband SEDs for each
light curve segment, the smoothness of these spectral breaks is impossible
to measure except in perhaps a few special cases.  In order to
learn about the global 
properties of X-ray afterglows in a statistical sample, we make simplifying
assumptions including neglecting the smoothness of these breaks.

We present an extensive set of closure relations (Table \ref{table:crs})
including those for constant 
density interstellar medium (ISM, $n(r)=$ constant) and wind ($n(r)\propto
r^{-2}$) environments, electron spectral index cases of $1<p<2$ and
$p>2$, and slow cooling and fast cooling regimes \markcite{sari98}({Sari} {et~al.} 1998).  For each case, we
present closure relations both with and without energy injection. These various
theoretical permutations have 
been invoked to explain the likely physical scenarios in the general scheme and
in many individual observational cases, but have never been combined on such a
large sample of afterglow light curves.

The constant density ISM environment, the simplest plausible model, is
often well fit by observations.  However, the wind environment is closer to
what one would expect for the medium surrounding
high-mass stars near the end of their lives (i.e. Wolf-Rayet stars).  This
dilemma has led to theoretical speculation of how massive star
environments might appear observationally to have a constant density.
\markcite{vanmarle06}{van Marle} {et~al.} (2006) conducted numerical simulations of stellar wind
environments where the wind exists only in a region very close to the star,
and showed that the right combination of wind pressure, ISM density,
progenitor rotation, and metallicity, can make a wind
environment appear like an ISM environment when probed by the GRB forward shock.
However, it is difficult to constrain these conditions observationally.
\markcite{chevalier07}{Chevalier} (2007) compares the environments and interactions of SNe Ib/c to the
expectations for the GRB case, finding the main difference to be in the
progenitor metallicity, ISM pressure, and possibly rotation.  He also
finds that GRB environments could produce a close 
termination shock outside of which the medium would be constant density.
Both the ISM and wind environments are surely simplistic descriptions of
actual GRB environments, especially when considering that local examples of
Wolf-Rayet stars like Eta Carinae show massive irregular winds that can be
modeled by neither a constant density nor a $r^{-2}$ environment.  In this
study, it is only feasible to test the simplest models, therefore we choose
to include both ISM and wind closure relation alternatives and determine
directly which model best approximates each afterglow.

The electron spectral index, $p$, is typically expected to be larger
than 2 to avoid a divergent total integrated energy for harder distributions
unless a high-energy 
cutoff is invoked.  Numerical modelling of electron acceleration
\markcite{achterberg01,eichler05}({Achterberg} {et~al.} 2001; {Eichler} \& {Waxman} 2005) is also consistent with $p>2$.  However, a $p<2$ flat
electron spectrum has been adapted to explain specific observations of
shallow temporal decays \markcite{panaitescu01,bhattacharya01}({Panaitescu} \& {Kumar} 2001; {Bhattacharya} 2001).  Therefore, we
choose to include all plausible alternatives.  When evaluating the closure
relations for $1<p<2$, we adopt the treatment described by \markcite{dai01}{Dai} \& {Cheng} (2001).

We evaluate sets of closure relations whose form is 
\begin{equation}\label{eq:cr}
  \alpha=f(\beta)
\end{equation}
by defining a new parameter, $\Psi$, such that:
\begin{equation}\label{eq:psi}
  \Psi \equiv \alpha_{obs}-f(\beta_{obs}).
\end{equation}
Using the measured
values of $\alpha_{obs}$ and $\beta_{obs}$, we compare the output ($\Psi$)
to the expectation of zero.  A relation 
is deemed valid if the output value is consistent within the error found by
propagating the $2\sigma=95.4\%$ confidence measurement errors on $\alpha$ and
$\beta$.

In the case of energy injection, the closure relations give
$\alpha=f(\beta,q)$, where $q$ is the luminosity index defined as
$L(t)=L_0(t/t_b)^{-q}$, which is physically valid for $q<1$
\markcite{zhang06,zhang01}({Zhang} {et~al.} 2006; {Zhang} \& {M{\'e}sz{\'a}ros} 2001).
The energy injection closure relations in Table \ref{table:crs} reduce to the
normal isotropic relations 
when $q=1$.
We solve for $q=g(\alpha,\beta)$ using the observed values
to determine $q$, as 
\begin{equation}\label{eq:q}
  q_{obs}=
  \begin{cases}
    g(\alpha_{obs},\beta_{obs}) &  (g<1) \\
    1 & (g \geq 1)
  \end{cases}
\end{equation}
We then calculate $\Psi$, as $\Psi=\alpha_{obs}-f(\beta_{obs},q_{obs})$.
Consequently when $q_{obs}<1$, by the criteria set in Equation \ref{eq:q},
$\Psi=0$. 

\subsection{Segment I} \label{sec:early}
The steep decline of segment I in the canonical X-ray light curve is probably
due to the tail end of the prompt emission and is governed by the curvature
effect, for which emission from different viewing angles reaches the
observer with different delays due to light propagation effects
\markcite{kumar00,zhang06}({Kumar} \& {Panaitescu} 2000; {Zhang} {et~al.} 2006).  The relationship between the temporal and spectral 
slopes of the high latitude emission is
\begin{equation}\label{eq:highlat}
  \alpha=2+\beta
\end{equation}
or in our formulation
\begin{equation}\label{eq:phi_highlat}
  \Psi=\alpha-2-\beta
\end{equation}
and is independent of any of the environmental or other parameters that
affect the closure relations for the external shock.  Therefore, we only use
this relation to help discriminate segment I from 
other segments and we do not use this relation to constrain any of the burst 
properties explored throughout the rest of this paper.  We find that $\sim 41\%$ of
the segments I in our sample are consistent with this relation, with the
remainder either steeper ($\sim 26\%$) or shallower ($\sim 34\%$) than this relation.

\markcite{zhang07a,zhang09}{Zhang} {et~al.} (2007b, 2009) explore steep declines in the XRT afterglow data set and
discuss several physical explanations, finding that steep decays with and
without significant spectral evolution can be explained by the curvature
effect.  They explore the 
subset with distinct spectral evolution and no clear and obvious
contaminating flaring to conclude that the data are best characterized by an
apparent evolution of a cutoff power-law spectrum.  Those that are
contaminated with flaring mix spectral evolution during the flares with
possible spectral evolution of the underlying afterglow, leading to
difficulty in characterizing the mean spectral and temporal properties.  
\markcite{zhang09}{Zhang} {et~al.} (2009) can interpret the apparent spectral
evolution during the steep decay phase using a curvature effect model that
invokes a non-power-law spectrum at the end of the prompt emission phase. 
Those afterglows with $\alpha$ steeper than
$2+\beta$ could also be caused by only seeing the tail
of a flare and interpreting it as the steep decay.  This is further complicated
by the choice of $t_0$ for determining the slope of the temporal decay that
affects segment I much more strongly than the other segments
\markcite{zhang06,liang06}({Zhang} {et~al.} 2006; {Liang} {et~al.} 2006). 
\markcite{zhang07a}{Zhang} {et~al.} (2007b) fit $t_0$ in their multi-component spectral evolution models
to best characterize the temporal decay, rather than using the BAT trigger
time.  This method could help explain some of the deviations.  However,
these efforts are beyond the scope of this study and have little consequence
for the analysis of segments II-IV and the study of jet breaks, so are
not repeated here. 

\subsection{Segments II-III}
The closure relations governing external shocks depend on the local
environment, range of electron spectral index, cooling regime, energy
injection, peak frequency and cooling frequency.  The pre-jet break
``isotropic'' relations and the post-jet break relations are given in Table
\ref{table:crs}.
The framework for the closure relations for the slow and fast cooling cases is 
explored by \markcite{sari98}{Sari} {et~al.} (1998) and expanded upon for the collimated case by
\markcite{sari99}{Sari} {et~al.} (1999).  \markcite{dai01}{Dai} \& {Cheng} (2001) include the $1<p<2$ cases, with additional
cases given in \markcite{zhang04}{Zhang} \& {M{\'e}sz{\'a}ros} (2004).  The addition of the energy 
injection mechanism to explain the plateau portion of X-ray light curves
brought additional parallel closure relation sets for $p>2$ given in
\markcite{zhang06}{Zhang} {et~al.} (2006).  We extract the additional jet break relations from
information provided in \markcite{panaitescu05b}{Panaitescu} (2005) and \markcite{panaitescu06a}{Panaitescu} {et~al.} (2006).  We
choose not to include the cases of energy injection for $1<p<2$ because this
scenario is unduly complicated, unlikely, and often not analytically solvable.

In segments II-III, the relationship between the electron slope ($p$) and
the measured X-ray spectral slope ($\Gamma$) is derived in \markcite{sari98}({Sari} {et~al.} 1998) as:
\begin{equation}\label{eq:beta}
  \Gamma-1 \equiv \beta=
  \begin{cases}
    1/2 & \nu_c<\nu<\nu_m \\ 
    (p-1)/2 & \nu_m<\nu<\nu_c \\
    p/2 & \nu_m,\nu_c<\nu \\
  \end{cases}
\end{equation}
where $\nu_m$ and $\nu_c$ are the synchrotron and cooling frequencies,
respectively.



\subsection{Segment IV}

Other large scale studies \markcite{liang08,panaitescu07,evans09}({Liang} {et~al.} 2008; {Panaitescu} 2007; {Evans} {et~al.} 2008) of jet break closure
relations address only the simplest cases of the uniform jet
\markcite{zhang04}({Zhang} \& {M{\'e}sz{\'a}ros} 2004). We also include the non-spreading uniform jet
with energy injection, the laterally spreading uniform jet with and
without energy injection and the simplest form of structured jets with
power-law angular distribution of energy outflow from
\markcite{panaitescu06a}{Panaitescu} {et~al.} (2006). We apply all of these cases for both  
ISM and wind environments.  We choose not to apply any of the $1<p<2$
closure relations that include energy injection and occur post-jet break, as
well as anything more complex than the most simple $p>2$ structured jet
relations due to their complexity and impracticability.  Closure relations
for all jet models are listed in Table \ref{table:crs}.  All of these
relations are for on-beam geometry.  In particular, the power-law structured
jet model requires that the line of sight is within the central cone beam.

The power-law structured jet relations are valid for a particular $\alpha$
and $\beta$ provided the index, $k$, of the angular energy distribution is less
than $\tilde{k}$, where 
\begin{equation}\label{eq:k}
  \tilde{k}=
  \begin{cases}
    \frac{8}{2\beta+5} \ (\textrm{ISM}\ \nu<\nu_c) \\
    \frac{8}{2\beta+3} \ (\textrm{ISM}\ \nu>\nu_c) \\
    \frac{8}{2\beta+4} \ (\textrm{Wind}) \\
  \end{cases}
\end{equation}
The case of  $k>\tilde{k}$ reduces to the non-spreading uniform jet relations 
because the core is dominant \markcite{panaitescu05b}({Panaitescu} 2005).


\subsection{Internal Consistency Checks}
The models in Table \ref{table:crs} can be used in succession throughout
an individual X-ray afterglow light curve, however several models often fit
equally well for any given light curve segment.  After fitting each closure
relation to each set of $\alpha$ and $\beta$, we combine them to form a coherent
physical model for each afterglow.  While we often cannot distinguish a unique
set of closure relations, we can use information from one segment to exclude
inconsistent relations from other segments.  
We assume there is no perceptible change in the circum-GRB environment
probed throughout an individual afterglow, and therefore if either an ISM or Wind
environment can be excluded in any segment, we exclude it for the 
other segments.  We do not allow transitions from slow cooling to fast
cooling within any light curve.  Since we do not see evidence of a change of
spectral index between segments in most individual afterglows, we require
the spectral regime to remain 
constant throughout a light curve (i.e. either $\nu<\nu_c$ or $\nu>\nu_c$ for
slow cooling, or $\nu_c<\nu<\nu_m$ or $\nu>\nu_m$ for fast cooling), therefore
excluding light curve breaks due to transitions of the cooling frequency
through the X-ray band.
Theoretically, we do not expect the electron spectral index, $p$, to change
during an afterglow light curve.  However, because we do not include the full
suite of $1<p<2$ relations with energy injection, we are careful not to
exclude a $1<p<2$ model just because only $p>2$ models are available to fit
other segments.  We eliminate models only on the the basis of their $p$ values
being inconsistent between segments, which is especially important for those
segments with $p\sim2$ and large error bars.
We require a segment consistent with a 
particular isotropic model to also be consistent with 
a corresponding jet model in the following segment, and conversely a jet model
must be consistent with a corresponding previous isotropic model in terms of
environment and spectral frequency regime.
Similarly, any energy injection model in segment II must
be consistent with the models for the following segment (if present) in terms
of environment and frequency regime. 
When a post-jet break relation with energy injection is applied, we
require $q$ to be consistent between pre- and post-jet break segments.
We also only allow jet breaks to directly precede potential segments
III or IV.  Ambiguous single power-laws have no restrictions on
available sets of closure relations.

Using these criteria, we eliminate some models to better constrain a
coherent picture for each afterglow. Unfortunately a unique set of models
is often still unattainable, as several equally likely relations remain.  In an
attempt to extract any available information from the remaining closure
relations, we look at constraints from families of closure relations to
determine specific properties.  For example, if several closure relations
are consistent with a particular segment or whole afterglow but they are all
ISM relations, then we can conclude that this afterglow is consistent with
the expectations of an ISM
environment. This also works for families of relations with a common
spectral frequency regime, environment, and pre/post-jet break regime.

\subsection{Classification}
Our sample is broken up into four groups (as described in
\S\ref{sec:lcfit}) based on their temporal properties.  These groups
include the I-II-III-IV/II-III-IV sample which have distinct segments
IV, segments I-II-III, ambiguous segments II-III/III-IV, and single
power-laws.  The latter three samples may contain jet breaks and
still fit within the canonical light curve picture.  The presence of
observing biases can explain the missing contextual clues that would make jet
break distinctions more clear.   In order to avoid imposing biases on the
results, we have not attempted to distinguish jet breaks 
based on {\it a priori} assumptions for decay indices or break times.  

The fact that segments II-III/III-IV light curves do not look like
the canonical light curve is likely a result of two different
observing biases: a late start or early end.  The behavior prior to
segment II was either not observed at all due to a late observation
start, or was complicated by flaring behavior making the underlying
afterglow shape unclear, both leading to an unknown initial steep
decline.  If the observations began even later, segment I and II would
be missed, leaving only segments III-IV.  Similarly, an early
observation end would also lead to the observer missing the jet
break.  This end could be due to either a manual end of the
observations because of observing contraints, or the afterglow
becoming too faint for XRT detection.
These two-segment ambiguous light curves may contain jet breaks as
either the observed break, or with the inclusion of energy injection, the jet
break would have occurred sometime prior to the start of the first segment.  In
this latter case, the actual jet break time is impossible to determine because
either there were no observation prior to the start of the apparent segment
II.  Considering the wide range of decay indices, observing times, and
redshifts, these limitations are likely to have influenced the data.

Apparent segments I-II-III show some suggestions of deviation from the
canonical form, which can also be explained by observing biases.
In some cases, the temporal
decay of these segments III is substantially steeper than typical segments
III, as if they transitioned directly from segment II to segment IV.
Alternatively, it is possible that the segment III is simply missing or buried
in the data due to a short duration, large error bars, or gaps in the light
curve with segment IV appearing to follow directly after segment II.
\markcite{willingale07}{Willingale} {et~al.} (2007) proposed an empirical form of the canonical light curve that
is made up of two falling exponential plus 
power-law functions that can explain the structure of most X-ray afterglow
light curves.  They suggest that missing phases, like that seen in the
apparent II-IV transitions, are a result of one of the
two components being particularly bright, weak, or short lived.

Other types of light curves exist in the sample for which we do not observe
jet breaks, and would not expect them to be hidden or ambiguous.
These include light curves showing only segments I-II, where the
later segments presumably occurred after observations ended.
Another type and possible jet break sample contaminant is the
so-called ``naked bursts''  
which show only the initial steep decline from the prompt emission without
the subsequent segments (II-IV) attributed to the afterglow.  This is
thought to occur because the surrounding medium is not dense enough to
produce the external shocks that power GRB afterglows \markcite{kumar00}({Kumar} \& {Panaitescu} 2000).
These GRBs can appear as either single power-law decays or broken power-laws
where the prompt emission and high latitude emission masquerade as segment
II-III light curves in the absence of comparison to the $\gamma$-ray prompt
emission.  Three examples of possible naked 
GRBs that have been investigated extensively are the long GRBs 050421
\markcite{godet06}({Godet} {et~al.} 2006) and 050412 \markcite{mineo07}({Mineo} {et~al.} 2007), and the short hard burst GRB
051210 \markcite{laparola06}({La Parola} {et~al.} 2006). 
To filter out these bursts we search for any light
curves that appear to be segments II-III, have break times $<1000$ sec,
and are consistent with the high latitude closure relation
(Eq.~\ref{eq:highlat}); these should not be considered
candidates for jet breaks.  Using these criteria, we identify GRB 060801 as
another possible naked burst in addition those identified in the
literature that is also a short hard GRB.  There may be more naked
bursts in the single power-law sample 
like that of GRB 050412, but we have no clean way of distinguishing them,
therefore we leave them for later discrimination.
It is possible that additional naked bursts are present in
the 29 GRBs excluded from the final sample due to their faintness and limited
observations which made temporal and spectral analysis not possible.

In order to learn more about these ambiguous afterglows and distinguish jet
breaks, we apply the closure relations.  The following section
describes how we use the closure relations and the temporal behavior
to identify additional jet break candidates.

\section{Jet Breaks} \label{sec:results}
Using the above criteria and consistency checks, we define sub-samples of
X-ray afterglows that potentially contain jet breaks.
Large errors on $\alpha$ and $\beta$ sometimes make definitive determination of
afterglow properties via closure relations unfeasible.  The cases where many
(pre- and post-jet break) closure relations are consistent are evaluated using
additional criteria and classified into categories based on their likelihood of
containing real jet breaks.  These additional criteria are based upon their
resemblance to those afterglows with clear canonical jet breaks.  We distinguish
between those afterglows which are consistent with {\it only} post-jet break
closure relations, those that are {\it only} consistent with pre-jet break
closure relations, and those that are consistent with both.
We divide our sample of potential jet breaks into 4 categories.  The
categories are the Prominent jet breaks, Hidden jet breaks, Possible jet
breaks, and Unlikely jet breaks.
The details of their categorical definitions follow.

\subsection{Prominent Jet Break Class} \label{sec:prominent}
\begin{figure}
  \begin{center}
    \includegraphics[scale=0.9]{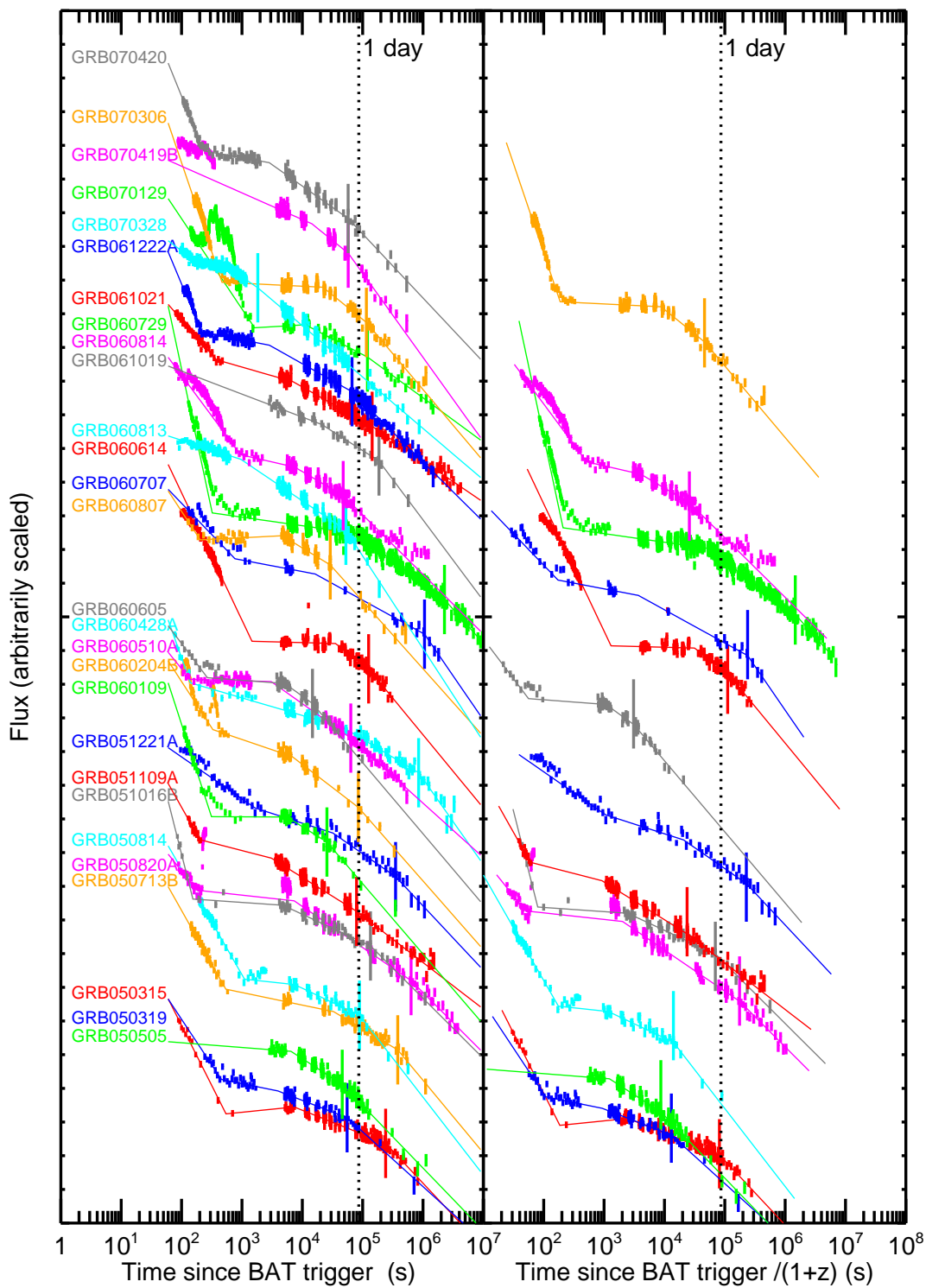}
  \end{center}
  \caption{Arbitrarily scaled light curves and temporal fits for all
    Prominent jet breaks in the observed frame 
    ({\it left}) and rest frame ({\it right}, where
    available). The final light curve break is indicated by the vertical line
    in the same color as the light curve and fit.  \label{fig:prom}} 
\end{figure}

We define the Prominent jet break class as those light curves for
which we can clearly distinguish a break between segments III and IV, with
segment IV being consistent with post-jet break closure relations,
consistent with the canonical morphology (Figure \ref{fig:prom}).  This
conservative classification criterion requires that 
the light curve is composed of either segments I-II-III-IV or II-III-IV, and
the final segment is consistent with post-jet break closure relations.
We find 30 such afterglows, 28 of which are consistent 
within $2\sigma$ of at least one post-jet break closure relation in their
segment IV after internal consistency checks.

In the two inconsistent cases (GRBs 061121 and 070508), all models
were eliminated in the process of internal consistency checks with the
$2\sigma$ criteria.  The unusually bright GRB 061121 \markcite{page07}({Page} {et~al.} 2007) was
triggered by a precursor, causing the choice of $T_0$ to affect the slopes,
which may account for some of the deviations from the models.  \markcite{page07}{Page} {et~al.} (2007)
also suggest the presence of a Comptonized component.
Therefore, these outlier cases may not be well represented by the canonical
form, have breaks due to other mechanisms such as the cooling frequency
moving through the X-ray band, or are not valid with the suite of 
closure relations used here, and are ignored in the following analysis.

\begin{figure}
  \begin{center}
    \includegraphics[scale=0.6]{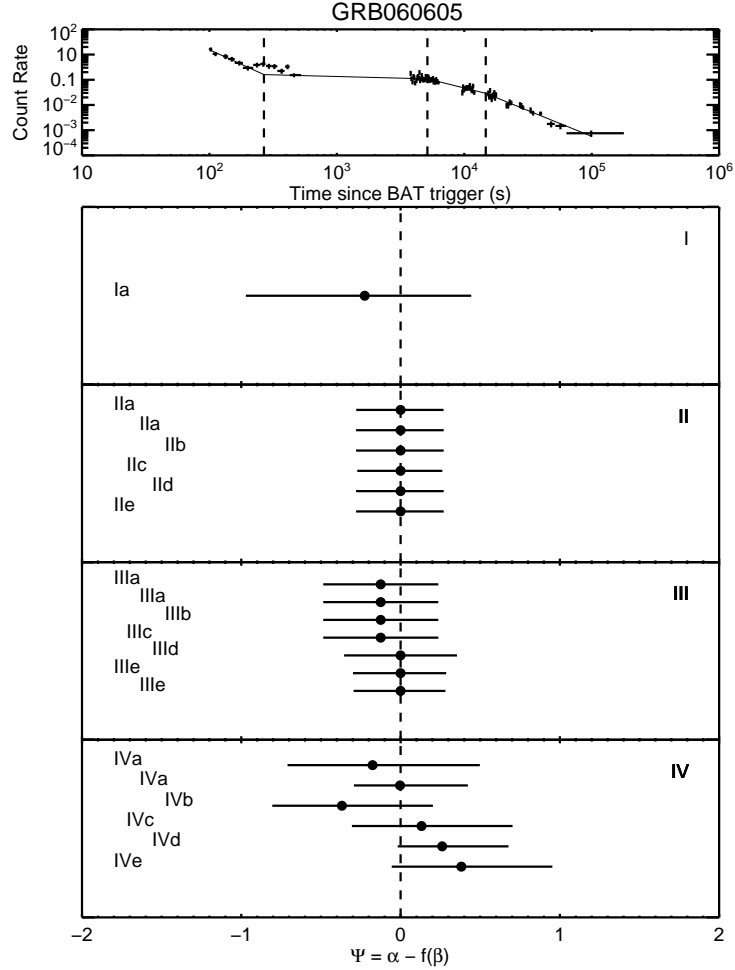}
  \end{center}
  \caption{Top panel shows the light curve in ($0.3-10.0$ keV counts $s^{-1}$)
    and fits for GRB 060605, an example of a Prominent Jet Break. 
    The four lower panels show the fits to the closure relations for
    the cases that are consistent with the data at the $2\sigma$ confidence level.
    An x-axis value consistent with $\Psi=0$ (dashed line) is valid.  The relations
    that involve energy injection require the $q$ parameter to be consistent
    with $q<1$, which is evaluated by the $\Psi$ parameter (Equation
    \ref{eq:psi}).  
    Ia - {\it HighLat}; 
    IIa - {\it ISMs2ai} ($p=3.20_{-0.43}^{+0.44},q=0.02_{-0.18}^{+0.17}$); IIb - {\it ISMs3ai} ($p=2.20_{-0.43}^{+0.44},q=0.03_{-0.27}^{+0.26}$); IIc - {\it WINDs3ai} ($p=2.20_{-0.43}^{+0.44},q=0.03_{-0.27}^{+0.26}$); IId - {\it ISMf2ai} ($q=0.07_{-0.60}^{+0.58}$); IIe - {\it ISMf3ai} ($p=2.20_{-0.43}^{+0.44},q=0.03_{-0.27}^{+0.26}$); IIf - {\it WINDf3ai} ($p=2.20_{-0.43}^{+0.44},q=0.03_{-0.27}^{+0.26}$); 
    IIIa - {\it ISMs3a} ($p=2.51_{-0.33}^{+0.36}$); IIIb - {\it WINDs3a} ($p=2.51_{-0.33}^{+0.36}$); IIIc - {\it ISMf3a} ($p=2.51_{-0.33}^{+0.36}$); IIId - {\it WINDf3a} ($p=2.51_{-0.33}^{+0.36}$); IIIe - {\it JETs3ai} ($p=2.51_{-0.33}^{+0.36},q=0.17_{-0.24}^{+0.24}$); IIIf - {\it JETsISM2ai} ($p=3.51_{-0.33}^{+0.36},q=0.27_{-0.17}^{+0.17}$); IIIg - {\it JETsISM3ai} ($p=2.51_{-0.33}^{+0.36},q=0.36_{-0.24}^{+0.23}$); 
    IVa - {\it JETs3a} ($p=2.23_{-0.46}^{+0.54}$); IVb - {\it JETs3b} ($p=2.23_{-0.46}^{+0.54}$); IVc - {\it JETsISM2a} ($p=3.23_{-0.46}^{+0.54}$); IVd - {\it JETsISM3a} ($p=2.23_{-0.46}^{+0.54}$); IVe - {\it JETsISM3b} ($p=2.23_{-0.46}^{+0.54}$); IVf - {\it JETsWIND3a} ($p=2.23_{-0.46}^{+0.54}$)
    \label{fig:gold_example}}
\end{figure}

Based on the canonical light curve form, we assume that segment IV is post-jet
break.  Therefore, the post-jet break closure relations are the only models
allowed in segment IV.  In contrast, many closure relations
are allowed by the canonical form for segments II and III.
Figure \ref{fig:gold_example} shows an example
of a burst in the Prominent jet break category with its light curve and
the closure relations that are consistent with the data.  All four segments
of this burst can be adequately characterized by the closure relations:
segment I is consistent with the high latitude relation; segment
II requires $p>2$ and energy injection, but is consistent with either
slow or fast cooling and either ISM or wind environment; segment
III is consistent with either isotropic or jet-with-energy-injection relations;
and segment IV is consistent with either a spreading or non-spreading
jet with an ISM or wind environment.  The resulting fits imply that the
jet break time cannot be unambiguously established; if the II-III break is
due to the cessation of energy injection then the III-IV break is a jet
break, but it is also possible that the II-III break is the jet break and
the III-IV break is due to the end of energy injection (see also
\markcite{ferrero09}{Ferrero} {et~al.} 2008 for a similar analysis of this GRB). 
We list the properties of the Prominent jet break bursts in Table
\ref{table:prominent}.  The distributions of these properties will be discussed
in \S\ref{sec:properties}.

We wish to assess the deficit of jet breaks in the XRT afterglow sample,
therefore we must 
make a reasonable estimate of the fraction of our sample with jet breaks,
accounting for a variety of observing biases.  Due to various observing
constraints and light curve profiles, not all burst observations began with an
immediate slew nor were they all observed out to a time at which the jet break is
expected to have occurred.  Therefore, to calculate an accurate jet break
fraction, we reduce our sample to only those GRBs
for which the observations span a time frame where we would expect a jet
break.  Previous studies of optical jet breaks \markcite{frail01,bloom03,zeh06}({Frail} {et~al.} 2001; {Bloom} {et~al.} 2003; {Zeh} {et~al.} 2006) showed
them to occur several days after the GRB trigger.  Instead of making
{\it a priori} assumptions about achromaticity or assuming similar behavior, we
determine the time frame during which we would expect a jet break by
studying the Prominent jet break sample.

\begin{figure}
  \begin{center}
    \includegraphics[scale=0.7]{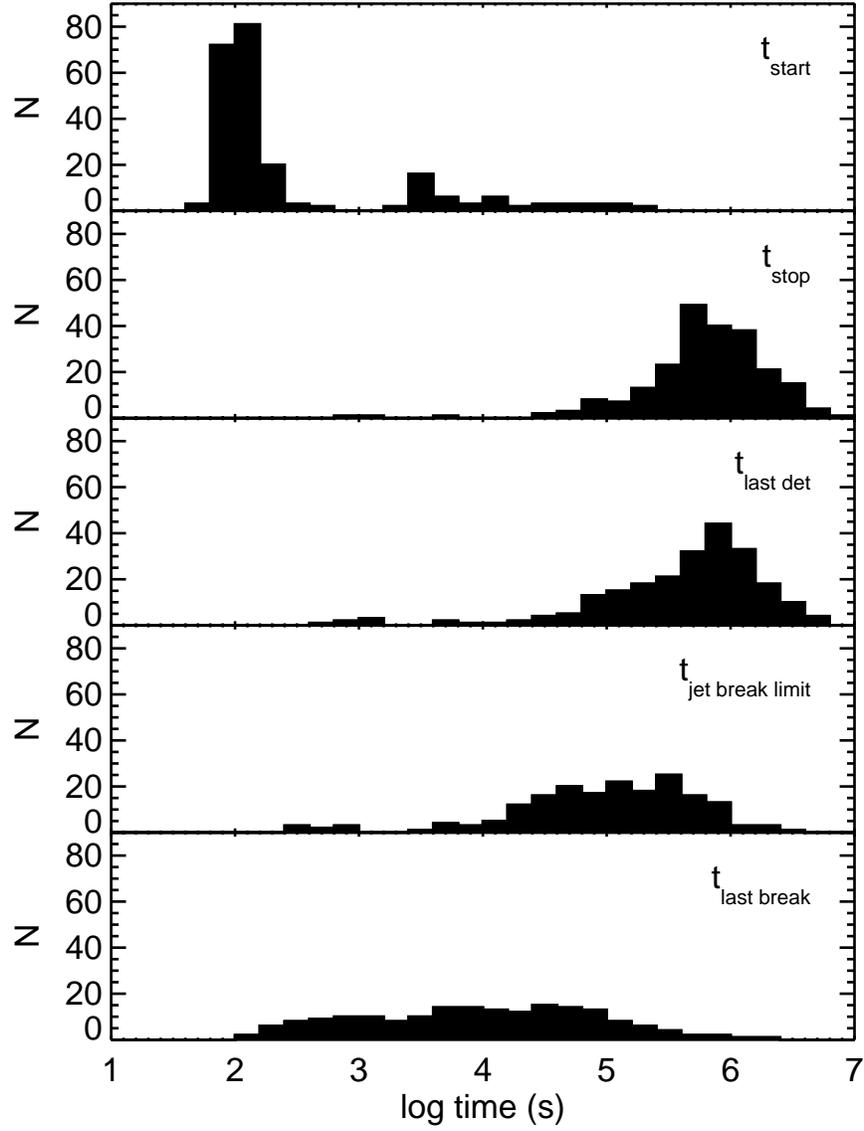}
  \end{center}
  \caption{Distributions of observation start times, stop times, time of
    last detection, jet break lower limit, and time of last measured
    breaks.
  \label{fig:times}}
\end{figure}

The Prominent jet break sample consists of 28 X-ray afterglows with $t_{start}$
ranging from a few minutes for those light curves that start with a segment I
to an hour for those that start during segment II, and $t_{stop}$ ranging
between 2 days and 5 months (excluding late-time {\it Chandra}
observations). The important measurement from the Prominent 
sample is the jet 
break time ($t_{b}$) which ranges between 0.02 and 26.2 days (excluding
earlier breaks that suggest post-jet-breaks-with-energy-injection). Excluding the
extremely late jet break case of GRB 060729, and the extremely early jet break
case of GRB070328, the latest light curve break in the whole sample
is $\sim 12$ days, with $90\%$ occurring within 10 days.  Therefore, we 
define our ``complete'' sample (those GRBs for which observations
sufficiently cover a time range where a jet break could have been measured), to
begin before 0.1 days and end after 10 days.  Of the 230 GRBs in our sample, 82
fit these completeness criteria.  The distributions of observation start, stop,
break, last detection, and the jet break lower limit (described in
\S\ref{sec:disc}) are shown in Figure \ref{fig:times}.

\subsection{Hidden Jet Break Class} \label{sec:hidden}
The Prominent jet break category constitutes only $\sim12\%$ of the total
sample.  We examined the remaining light curves for evidence of
``hidden'' jet breaks, identified by their closure relations rather
than the light curve morphology.  Our Hidden jet break category
includes those light curves with ambiguous final segments.
We consider three ambiguous cases. As noted in \S\ref{sec:lcfit},
broken power-laws with $\alpha_2>\alpha_1$ cannot be distinguished
{\it a priori} between segments II-III and III-IV.  If the final
segment is only consistent with post-jet break closure relations, then
we designate it as a III-IV case with a Hidden jet break.  The second
ambiguous case involves three segment light curves initially
classified as I-II-III in which the final segment is steeper than
typical segments III and consequently only consistent with post-jet
break closure relations.  Therefore, this type appears more like a
I-II-IV or a jet-break-with-energy-injection.  The third ambiguous case
involves single power-law light 
curves which are only consistent with post-jet break closure
relations even though the jet break itself is not observed.

\begin{figure}
  \begin{center}
    \includegraphics[scale=0.6]{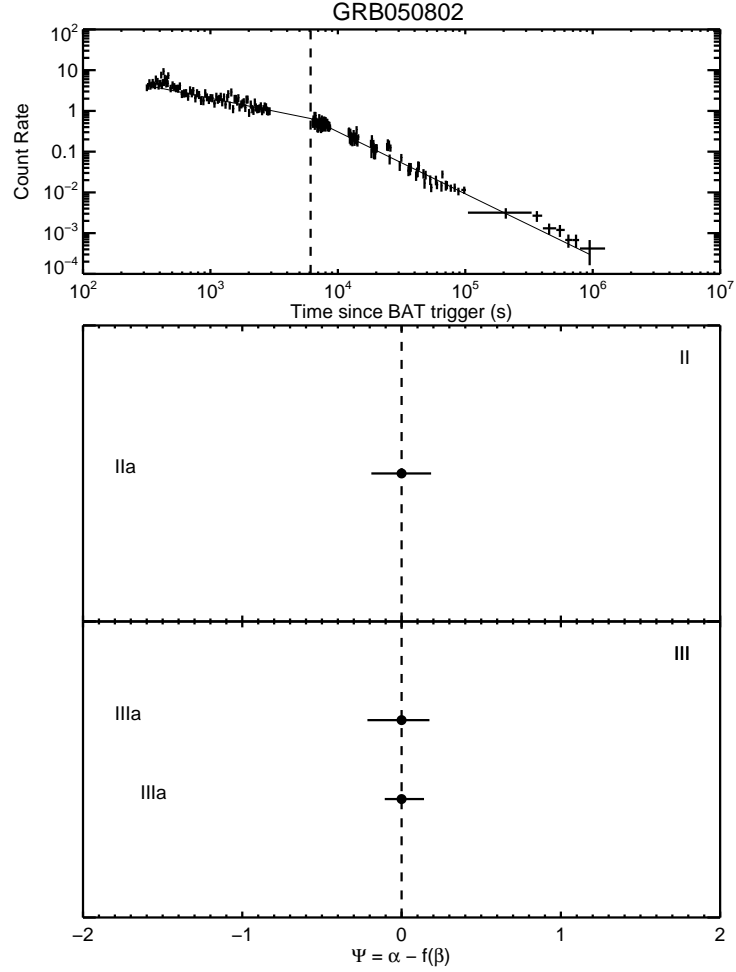}
  \end{center}
  \caption{Same as Figure \ref{fig:gold_example}, except for GRB
    050802, which is an example of the Hidden Jet Break
    category. 
    IIa - {\it ISMs2ai} ($p=2.73_{-0.24}^{+0.25},q=0.53_{-0.13}^{+0.13}$); 
    IIIa - {\it JETs2ai} ($p=2.71_{-0.26}^{+0.14},q=0.38_{-0.11}^{+0.09}$); IIIb - {\it JETsISM2ai} ($p=2.71_{-0.26}^{+0.14},q=0.70_{-0.11}^{+0.10}$)
    \label{fig:silver_example}} 
\end{figure}

We find an additional 12 light curves that fit these criteria, of
which 3 are from the 
two-segment ambiguous sample, 9 are segments III from the I-II-III sample, and
none are single power-laws.  Figure \ref{fig:silver_example} shows an example
that was classified as an ambiguous II-III/III-IV until we
found that its final segment is only consistent with post-jet break closure
relations.  We list the properties of the Hidden jet breaks in Table
\ref{table:hidden}.

Our classification of XRF 060218 as a Hidden jet break illustrates one
limitation of our methodology.  Using the closure relations
in Table \ref{table:crs}, we find a post-jet break decay to be the 
only possible outcome.  In doing so we assume that the emission is due
to a purely forward shock origin.  Individual studies on XRF/GRB 060218
\markcite{campana06,ghisellini07b}({Campana} {et~al.} 2006; {Ghisellini}, {Ghirlanda},  \& {Tavecchio} 2007b) show a strong early thermal component
related to the associated SN 2006aj and the observed shock break out,
followed by a possible Compton component.  Our analysis reveals an unusually
steep late-time spectral slope, which may suggest that our models are not
applicable in this individual case.  Therefore, this break may not be due to
a jet break at all.  We choose treat all GRBs in our sample in the same way, and
our methodology did not account for possible non-power law spectral
components.  As a result, a small fraction of the sample coule be
mis-classified.  We note that only $\sim 1\%$ of GRB afterglows show
evidence for thermal spectral components.

\subsection{Possible and Unlikely Jet Break Classes} \label{sec:feasible}
We now examine the remaining light curves to search for additional jet
breaks in the data set that have not been previously identified as
Prominent or Hidden.  Of the remaining 187 light curves, there are 43 that
are consistent with only pre-jet break relations in their last segment, and
therefore classified as non-jet breaks.  The remaining 144 light curves 
are ambiguous because they are consistent with both pre- and post-jet break
closure relations. 
This sample includes those ambiguous segments II-III/III-IV, single
power-laws, and segments I-II-III in which segment III is potentially
post-jet break.  We use the properties of the Prominent jet break
sample to identify possible jet breaks in the ambiguous sample.

\begin{figure}
  \begin{center}
    \includegraphics[scale=0.8]{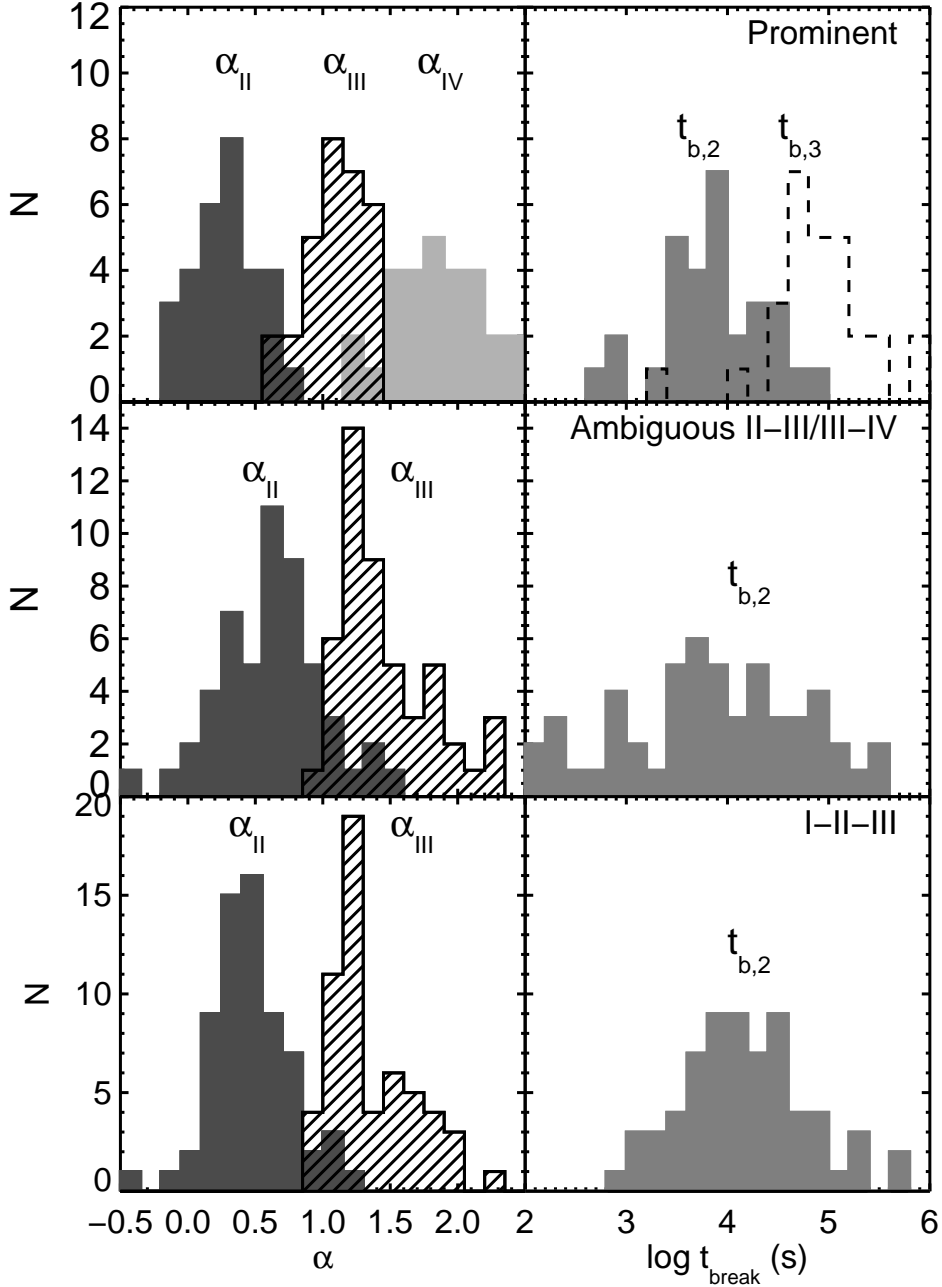}
  \end{center}
  \caption{Histograms of temporal decay indices ({\it left}) comparing segments
    II-III of Prominent jet break sample to those of the ambiguous
    II-III/III-IV sample and the segments I-II-III sample.  
    The distributions of break times ({\it right}) between segments II and III
    ($t_{b,2}$, grey) for all three samples are plotted with the break times
    between segments III and IV ($t_{b,3}$, dashed line) shown for comparison.  These
    distributions are similar with an excess at 
    larger break times in the ambiguous segments II-III/III-IV and segments
    I-II-III samples.
    \label{fig:compare23}}
\end{figure}

The samples of ambiguous segments II-III/III-IV, segments I-II-III, and the
Prominent sample have the common feature of apparent segments II-III with a
break in between, which is a shared distinct component that can be used to
compare them.  We use the Prominent jet break sample as a
``control group'' for comparisons with the other categories.
Figure \ref{fig:compare23} shows the 
distributions of temporal decay indices and break times for segments
II and III for all three groups.  The temporal indices split relatively
cleanly into several distinct distributions, especially in the
Prominent sample.  
There is a small amount of overlap between segments III 
and IV in the Prominent jet break sample, but this is not surprising
considering the multitude of possible model scenarios employed to explain
these light curves.  The distributions for the ambiguous segments
II-III/III-IV are plotted assuming that they are all segments II-III.
The broader and steeper distributions in this sample are consistent
with some contamination by actual segments III-IV.
The steeper than expected
segments II and III from the sample of segments I-II-III are 
also consistent with the hypothesis of segment III confusion and
post-jet-break-with-energy-injection.  The break times between
segments II and III, as plotted in the right-hand side of Figure
\ref{fig:compare23}, are also suggestive of these findings.  The distributions
are not nearly as narrow as those suggested by the canonical light curve form
\markcite{nousek06,zhang06}({Nousek} {et~al.} 2006; {Zhang} {et~al.} 2006), but 
do suggest a spread to larger break times in the contaminated ambiguous
II-III/III-IV sample and the segments I-II-III sample.  The break times are
also dispersed due to redshift effects whose amplitude is unknown for $\sim 60\%$
of the GRBs. When looking only at those GRBs with known redshifts, these same
distributions are narrower and more cleanly separated.

\begin{figure}
  \includegraphics[scale=0.5]{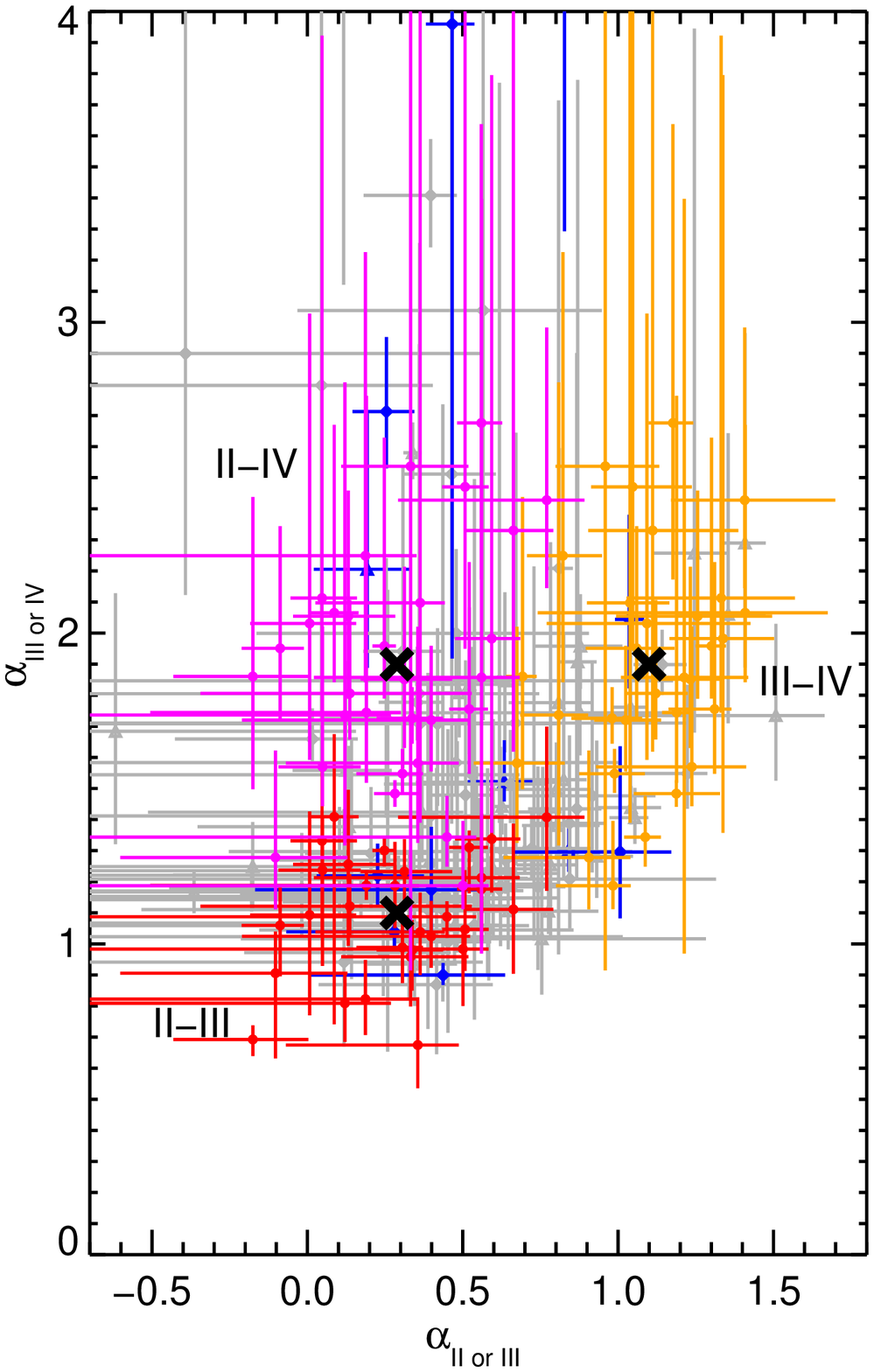}
  \includegraphics[scale=0.5]{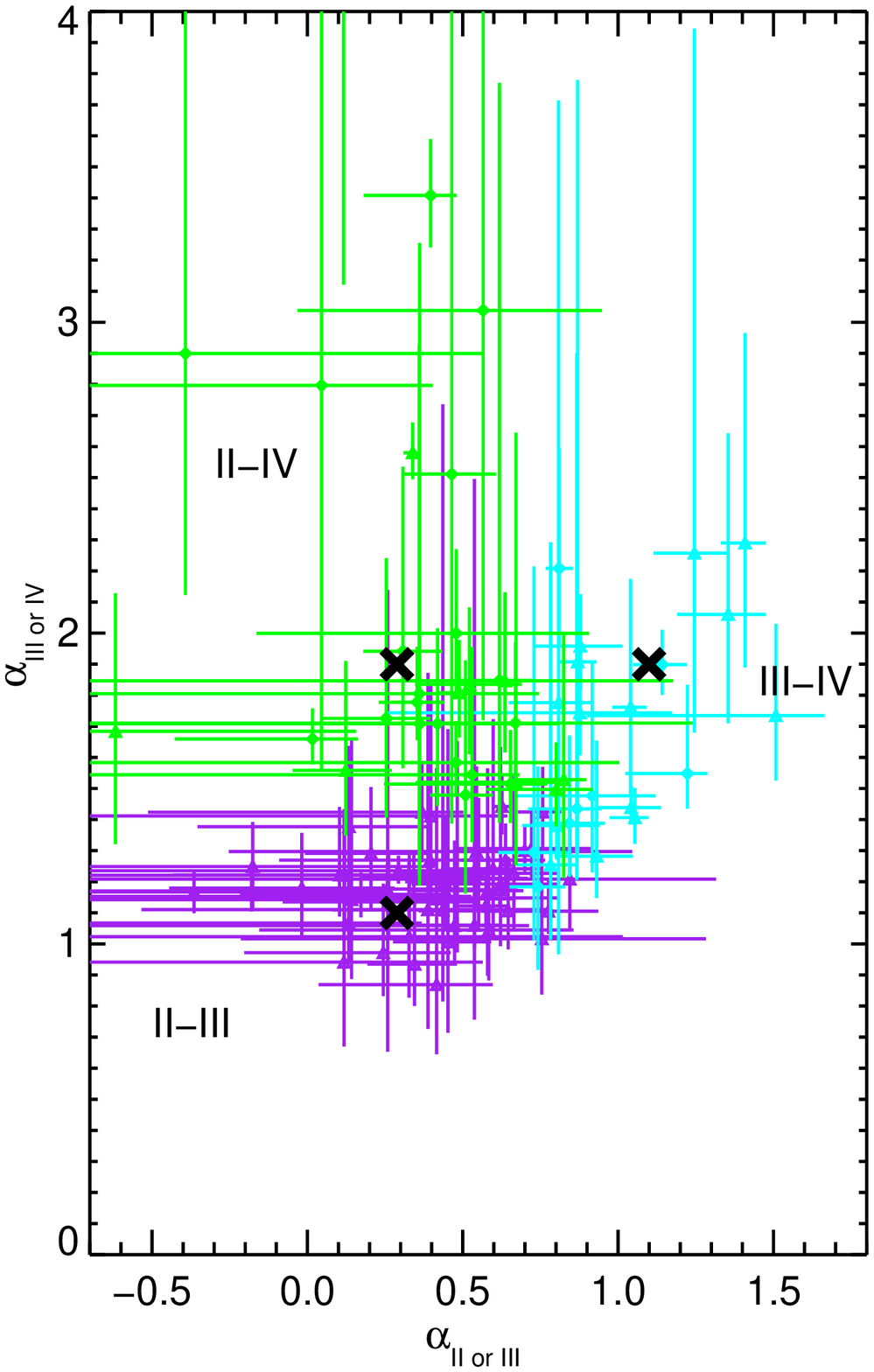}
  \caption{Correlation between temporal decays of Prominent jet
    break sample ({\it left}) segments II, III, and IV showing
    parameter space of segment 
    transitions II-III (red), III-IV (orange), and II-IV (magenta) used to
    classify the ambiguous transitions of the Hidden jet break sample
    (blue) and distinguish the Possible and Unlikely jet break samples
    (grey).  The black crosses mark the means of each potential transition group.
    The resulting classified transitions ({\it right}) are based upon
    their scaled proximity to the mean from the Prominent jet break
    sample with the newly classified segments II-III (purple), III-IV
    (cyan), and II-IV (green). Parameter errors are plotted with $2\sigma$
    confidence intervals. \label{fig:alpha_alpha}}
\end{figure}

To further distinguish those potential jet breaks that are in this contaminated
ambiguous group, we look at the correlations between the temporal decay slopes
from different segments in the same light curves.  In the left side of Figure
\ref{fig:alpha_alpha} we plot 
$\alpha_{III}$ versus $\alpha_{II}$, $\alpha_{IV}$ versus $\alpha_{III}$, and
$\alpha_{IV}$ versus $\alpha_{II}$  for the Prominent sample as our ``control
group'' and see a reasonably clean distinction between the light curve 
transitions in this $\alpha-\alpha$ parameter space, which can be used
to classify the remaining ambiguous segments.  
We use the ``control group'' to determine the area of $\alpha-\alpha$ space 
for each cluster of specific segment combinations.  The mean values
for each cluster from the control group are indicated by the black
crosses.  For each ambiguous light curve, we calculate the distance in
$\alpha-\alpha$ space to each cluster mean weighted by $\sigma_{\alpha}$, and 
categorize the ambiguous segment transitions based upon their proximity to
these mean values (right side, Figure \ref{fig:alpha_alpha}).

\begin{figure}
  \begin{center}
    \includegraphics[scale=0.6]{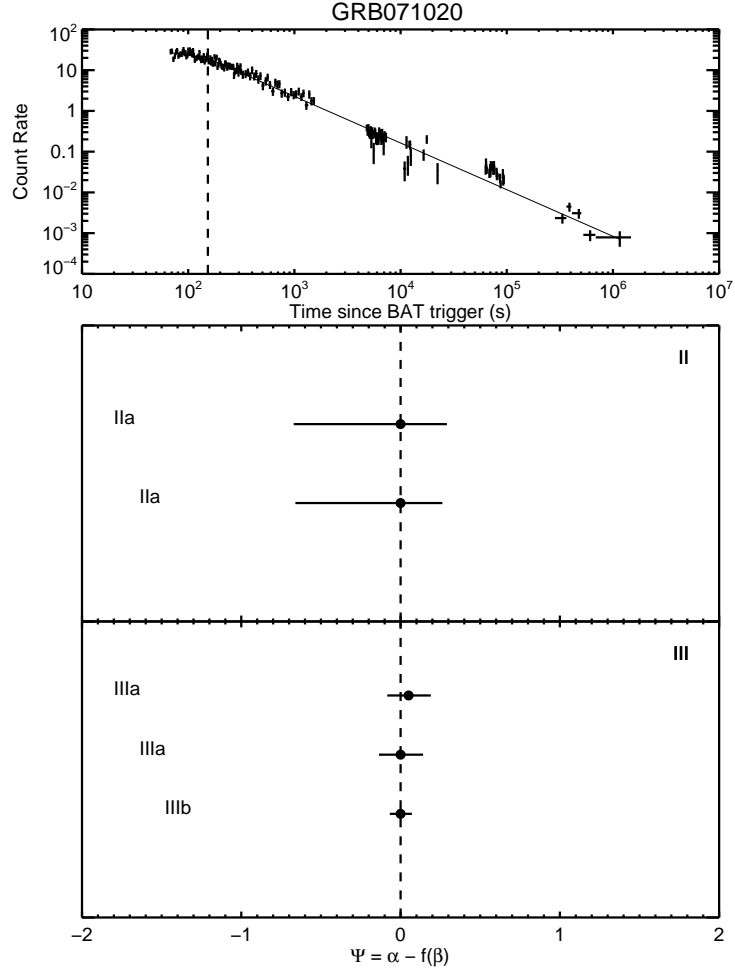}
  \end{center}
  \caption{Same as Figure \ref{fig:gold_example}, except that GRB 071020 is an
    example of those that were classified based upon the $\alpha-\alpha$
    parameter space criteria into an Unlikely jet break because it
    showed an apparent segment II-III transition. 
    IIa - {\it ISMs2ai} ($p=2.79_{-0.22}^{+0.24},q=0.36_{-0.46}^{+0.20}$); IIb - {\it ISMf2ai} ($q=0.94_{-1.19}^{+0.47}$); 
    IIIa - {\it ISMs2a} ($p=2.46_{-0.18}^{+0.18}$); IIIb - {\it JETs2ai} ($p=2.46_{-0.18}^{+0.18},q=0.28_{-0.07}^{+0.08}$); IIIc - {\it JETsISM2ai} ($p=2.46_{-0.18}^{+0.18},q=0.57_{-0.07}^{+0.07}$)
    \label{fig:possible_example23}}
\end{figure}

\begin{figure}
  \begin{center}
    \includegraphics[scale=0.6]{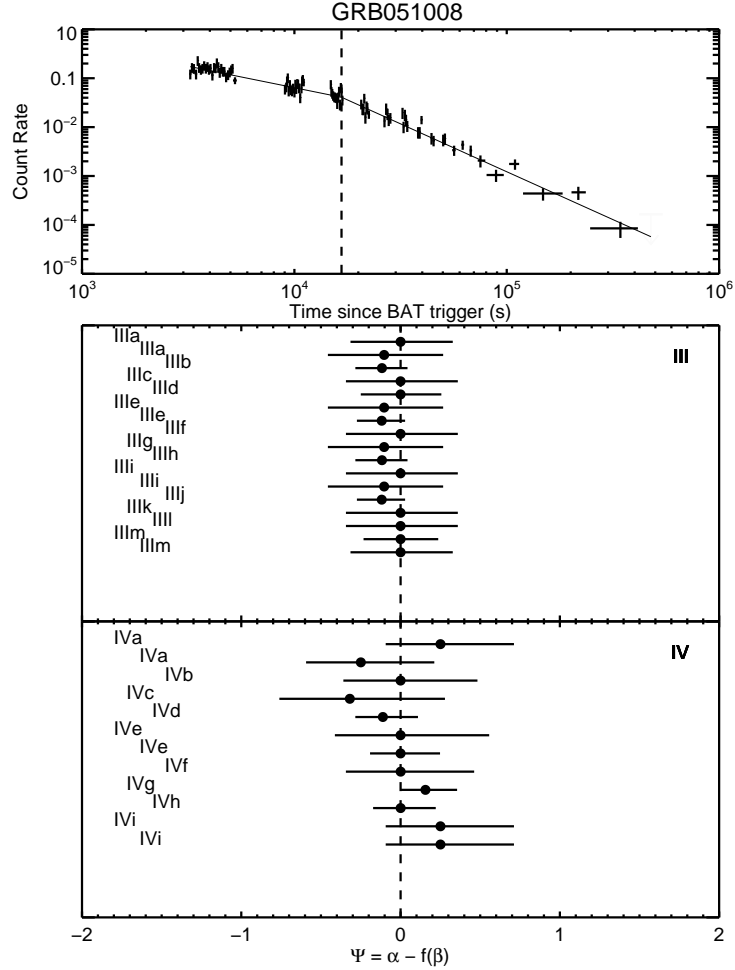}
  \end{center}
  \caption{Same as Figure \ref{fig:gold_example}, except that GRB 051008 is an
    example of those that were classified based upon the $\alpha-\alpha$
    parameter space criteria into a Possible jet break with an apparent segment
    III-IV transition.     
    IIIa - {\it ISMs2ai} ($p=2.98_{-0.43}^{+0.46},q=0.60_{-0.21}^{+0.22}$); IIIb - {\it ISMs3a} ($p=1.98_{-0.43}^{+0.46}$); IIIc - {\it ISMs3b} ($p=1.98_{-0.43}^{+0.46}$); IIId - {\it ISMs3ai} ($p=1.98_{-0.43}^{+0.46},q=0.90_{-0.35}^{+0.36}$); IIIe - {\it WINDs2ai} ($p=2.98_{-0.43}^{+0.46},q=-0.11_{-0.25}^{+0.26}$); IIIf - {\it WINDs3a} ($p=1.98_{-0.43}^{+0.46}$); IIIg - {\it WINDs3b} ($p=1.98_{-0.43}^{+0.46}$); IIIh - {\it WINDs3ai} ($p=1.98_{-0.43}^{+0.46},q=0.90_{-0.35}^{+0.36}$); IIIi - {\it ISMf3a} ($p=1.98_{-0.43}^{+0.46}$); IIIj - {\it ISMf3b} ($p=1.98_{-0.43}^{+0.46}$); IIIk - {\it ISMf3ai} ($p=1.98_{-0.43}^{+0.46},q=0.90_{-0.35}^{+0.36}$); IIIl - {\it WINDf3a} ($p=1.98_{-0.43}^{+0.46}$); IIIm - {\it WINDf3b} ($p=1.98_{-0.43}^{+0.46}$); IIIn - {\it WINDf3ai} ($p=1.98_{-0.43}^{+0.46},q=0.90_{-0.35}^{+0.36}$); IIIo - {\it JETs3ai} ($p=1.98_{-0.43}^{+0.46},q=0.17_{-0.26}^{+0.27}$); IIIp - {\it JETsISM3ai} ($p=1.98_{-0.43}^{+0.46},q=0.31_{-0.23}^{+0.24}$); IIIq - {\it JETsWIND3ai} ($p=1.98_{-0.43}^{+0.46},q=0.60_{-0.21}^{+0.22}$); 
    IVa - {\it ISMs2a} ($p=3.28_{-0.42}^{+0.57}$); IVb - {\it WINDs2a} ($p=3.28_{-0.42}^{+0.57}$); IVc - {\it JETs2ai} ($p=3.28_{-0.42}^{+0.57},q=0.37_{-0.17}^{+0.23}$); IVd - {\it JETs3a} ($p=2.28_{-0.42}^{+0.57}$); IVe - {\it JETs3b} ($p=2.28_{-0.42}^{+0.57}$); IVf - {\it JETs3ai} ($p=2.28_{-0.42}^{+0.57},q=0.78_{-0.29}^{+0.39}$); IVg - {\it JETsISM2ai} ($p=3.28_{-0.42}^{+0.57},q=0.73_{-0.17}^{+0.23}$); IVh - {\it JETsISM3a} ($p=2.28_{-0.42}^{+0.57}$); IVi - {\it JETsISM3b} ($p=2.28_{-0.42}^{+0.57}$); IVj - {\it JETsISM3ai} ($p=2.28_{-0.42}^{+0.57},q=1.00_{-0.26}^{+0.35}$); IVk - {\it JETsWIND3a} ($p=2.28_{-0.42}^{+0.57}$); IVl - {\it JETsWIND3ai} ($p=2.28_{-0.42}^{+0.57},q=1.16_{-0.23}^{+0.31}$)
    \label{fig:possible_example34}}
\end{figure}

\begin{figure}
  \begin{center}
    \includegraphics[scale=0.6]{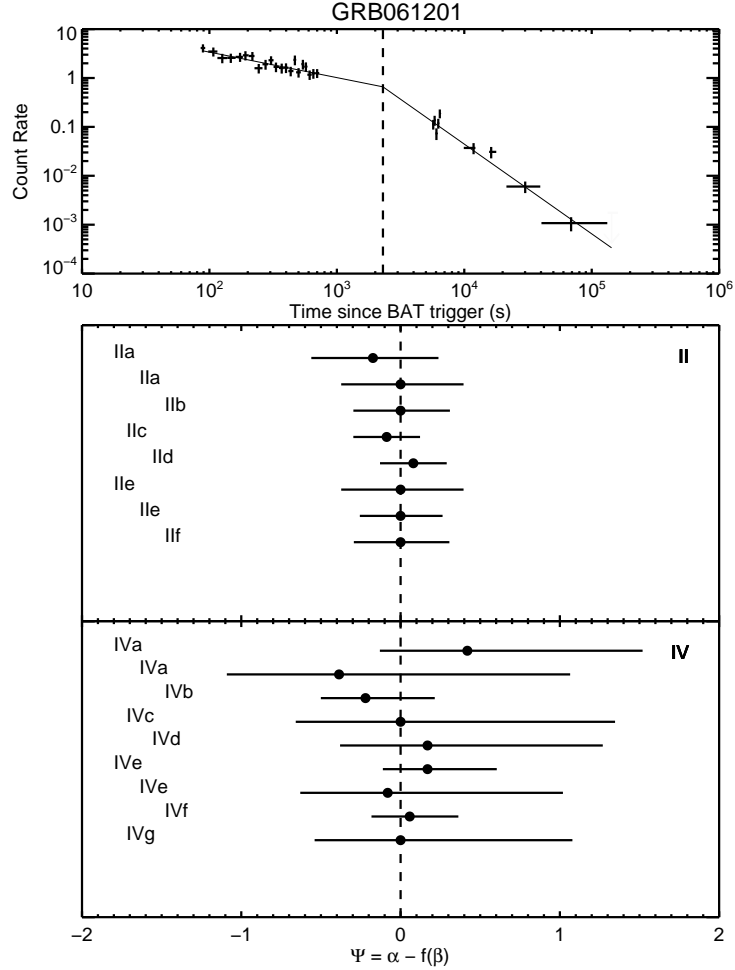}
  \end{center}
  \caption{Same as Figure \ref{fig:gold_example}, except that GRB 061201 is an
    example of those that were classified based upon the $\alpha-\alpha$
    parameter space criteria into a Possible jet break with an apparent segment
    II-IV transition.  
    IIa - {\it ISMs2a} ($p=1.93_{-0.46}^{+0.50}$); IIb - {\it ISMs2ai} ($p=1.93_{-0.46}^{+0.50},q=0.86_{-0.30}^{+0.32}$); IIc - {\it WINDs2ai} ($p=1.93_{-0.46}^{+0.50},q=0.08_{-0.41}^{+0.42}$); IId - {\it ISMf2ai} ($q=1.38_{-0.25}^{+0.24}$); IIe - {\it WINDf2ai} ($q=1.35_{-0.26}^{+0.26}$); IIf - {\it JETs2ai} ($p=1.93_{-0.46}^{+0.50},q=0.14_{-0.23}^{+0.24}$); IIg - {\it JETsISM2ai} ($p=1.93_{-0.46}^{+0.50},q=0.38_{-0.22}^{+0.23}$); IIh - {\it JETsWIND2ai} ($p=1.93_{-0.46}^{+0.50},q=0.05_{-0.24}^{+0.25}$); 
    IVa - {\it WINDs2a} ($p=2.22_{-0.67}^{+1.43}$); IVb - {\it JETs2a} ($p=2.22_{-0.67}^{+1.43}$); IVc - {\it JETs2b} ($p=2.22_{-0.67}^{+1.43}$); IVd - {\it JETs2ai} ($p=2.22_{-0.67}^{+1.43},q=0.78_{-0.38}^{+0.77}$); IVe - {\it JETsISM2a} ($p=2.22_{-0.67}^{+1.43}$); IVf - {\it JETsISM2ai} ($p=2.22_{-0.67}^{+1.43},q=1.11_{-0.36}^{+0.73}$); IVg - {\it JETsWIND2a} ($p=2.22_{-0.67}^{+1.43}$); IVh - {\it JETsWIND2b} ($p=2.22_{-0.67}^{+1.43}$); IVi - {\it JETsWIND2ai} ($p=2.22_{-0.67}^{+1.43},q=0.94_{-0.41}^{+0.83}$)
    \label{fig:possible_example24}}
\end{figure}

These resulting new classifications lead to 21 bursts previously in
the ambiguous 
segments II-III/III-IV or segments I-II-III sample that are similar in
$\alpha-\alpha$ space to the Prominent jet break sample for the segments III-IV
transition, and 26 bursts that are similar to the segments II-IV transition.
Their consistency with post-jet break closure relations and temporal slopes
suggest that there are indications of jet breaks in those light
curves.  
The remaining 53 ambiguous light curves from this sample's
transitions appear to be segments II-III and therefore probably pre-jet break.
However, they could possibly be segment II with a
post-jet-break-with-energy-injection segment III.  We are unable to
distinguish these cases, therefore we 
put these remaining afterglows into a new category called Unlikely jet
breaks. (These light curves are called Unlikely jet breaks because
their slopes are too shallow compared to the Prominent jet break
$\alpha$s to be post-jet break.)  
We show example light curves for each of these newly segregated groups in Figures
\ref{fig:possible_example23}-\ref{fig:possible_example24}.

Finally we examine single power-laws that are consistent with both
pre- and post-jet break closure relations.  Out
of the 48 afterglows in our sample that are fit by a single power-law,
none of the afterglows are solely consistent with post-jet break and not
pre-jet break relations.  
These single power-law light curves do not easily fit into the canonical
picture unless they are either a short snapshot of one segment or are the
post-jet break component and therefore relevant to this study.  Often 
these light curves are plagued by a low signal-to-noise ratio and few
counts, which leads to minimal information to be extracted. Other light
curves in this group are dominated by large flaring during part or all of
their light curves, which therefore makes determination of the underlying
afterglow shape impossible during that time interval.  We fit only the
portion of the light curves that clearly returns to the underlying
non-flaring level.  However, there are a few examples of outliers that do
not have flares and do have strong counting statistics.  
The most notable and best sampled X-ray afterglow in this outlier group is
that of GRB 061007 which displayed a very bright 
exceptionally smooth single power-law.  \markcite{schady07}{Schady} {et~al.} (2007) showed that this must
be either due to a very late-time jet break requiring enormous kinetic energy
or an exceptionally early jet break ($t_b<80\ s$) with highly collimated
outflow from a jet that includes continuous energy injection throughout.  We do
not exclude these isotropic models with extreme energy requirements from our
global study, but these extreme energy requirements are a valid
concern addressed in \S\ref{sec:energ}.  

\begin{figure}
  \begin{center}
    \includegraphics[scale=0.5,angle=90]{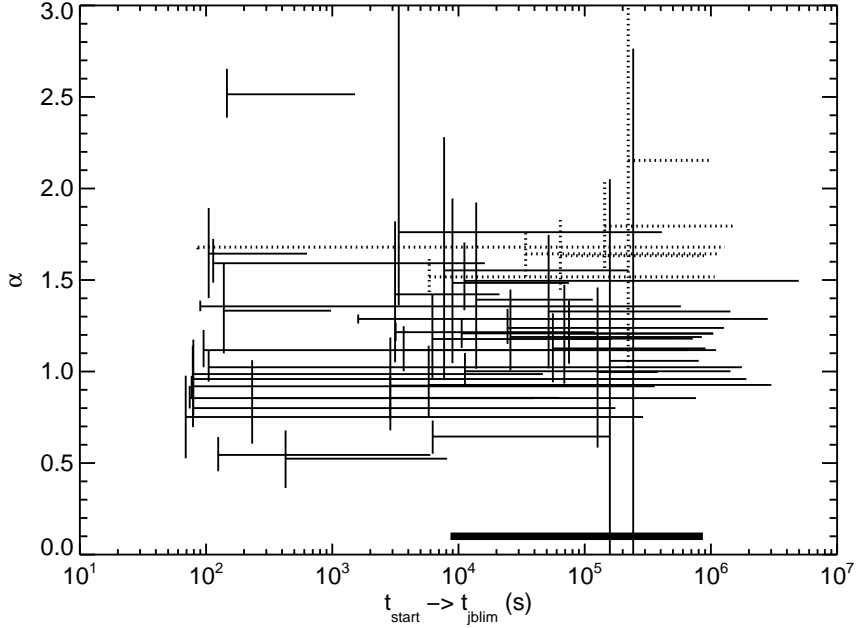}
  \end{center}
  \caption{Representation of those light curves that are best
    fit by a single power-laws and their duration from the start of
    observations to their last detection and the slope of the temporal decay.
    Note that these times exclude segments of flaring prior to or after the
    single power-law observations for which the shape of the underlying
    afterglow cannot be determined. The data suggest that the
    majority of light curves with shallower temporal  
    decays begin early and end early and steeper decays begin later and end 
    later. Note that these times are in the observed frame due to a lack
    of available redshifts for this sample.  The thick reference 
    line indicates the time interval during which jet breaks occur in
    the Prominent sample.  Those 
    bursts whose $t_{start}$ begins during the jet break time interval and
    whose $\alpha > 1.5$ are suspected to be jet breaks and indicated by the
    dotted lines.
    \label{fig:singlepl}}
\end{figure}

We attempt to filter the single power-laws that are a result of a short pre-jet
break segment from those that represent a post-jet break
decay.  In the context of the canonical X-ray afterglow, we should be
able to look at the relationship between observation start and stop times,
and the temporal decay to distinguish pre- and post-jet break.
Figure \ref{fig:singlepl} shows the relationship between the time that the
observations begin and the time of last detection as a function of $\alpha$.
Unfortunately there are only 14 redshifts measured of the 48 total GRBs in this
category that are consistent with at least one post-jet break closure relation.
Therefore, to include as many potential jet breaks as possible, the times used
in the following distinctions are in the observed frame and not the rest frame.
There appears to be a general trend that suggests that those single power-laws
for which observations began late ($>10^4$ s) tend to be steeper
($\alpha>1.2$) than those that start early and continue for a long observation.
We make a cut in this parameter
space and deem those afterglows that start within or traverse the time frame
for which we would expect jet breaks (from the Prominent sample) and have
steep ($\alpha > 1.5$) decays as 
Possible jet breaks and add them to that sample.  These 6 bursts are indicated in
Figure \ref{fig:singlepl}.  The remaining single power-law afterglows that appear
to be pre-jet break are put into the Unlikely jet break category.

We present the complete sample of Possible jet breaks in Table
\ref{table:possible}.  The probable jet break time for each light
curve is identified as $t_b$.  However, some ambiguity still remains of
whether this is the time of the jet break.  Therefore, we also list the
$t_{start}$ and $t_{lastdet}$ (the time of last detection, if relevant) 
because they provide limits on the jet break time if the jet break is not
$t_b$. Using the completeness criteria described in \S\ref{sec:prominent},
we find that at least $23\%$ of the Possible jet breaks and $35\%$ of
the Unlikely jet breaks were observed sufficiently long enough that we would
expect to have seen a jet break during their observations. 

The 6 afterglows that are inconsistent with all closure relations after
internal consistency checks, but have temporal behavior of their final
segments similar to Prominent segments III-IV or II-IV are also included in
the Unlikely jet break sample.  The breaks in these cases may be due to 
origins other than those in the canonical model such 
as transition of the cooling frequency through the X-ray band or Compton
processes.  They may also be contaminated by small scale flaring that is not
removed by our methods.  They are denoted in Table \ref{table:possible} by
``none'' in the requirements field.

The criteria, inputs, and final memberships of the Prominent, Hidden,
Possible, Unlikely, and Non-jet break categories are summarized in Table
\ref{table:summary}.  The non-jet break category includes all
remaining bursts, most of which are segment I-II transitions.

\subsection{Significance of Closure Relation Distinctions} \label{sec:properties}

We evaluated the statistical significance of the jet breaks identified
by closure relation distinctions by running a series of Monte Carlo simulations.
These specific samples include those GRBs for which we have a distinct segment
IV, ambiguous segment II-III, segment I-II-III with no
segment IV, and single power-laws.  
We generated 1000 mock sets of $\alpha$,
$\beta$, $\sigma_{\alpha}$, and $\sigma_{\beta}$ for each segment in each
simulation by drawing random numbers from the Gaussian distribution of each
of these parameters from the real data.  These 1000 sets of light
curve parameters for each sample are fit with the closure relations as
we did with the real data including the application 
of the internal consistency checks.  

The Monte Carlo simulations are used to determine how many
Hidden jet breaks might result from random variations in $\alpha$s
and $\beta$s due to measurement errors.
The fraction of Hidden jet breaks in this randomized sample indicates
to us how many false Hidden jet breaks we would expect to see in the
real sample. Based on the simulations, we would have expected $6.0 \pm 0.6$
final segments only consistent with post-jet break closure relations,
compared to the 12 found in the real data. Therefore, at least half of
these Hidden jet breaks appear to be real.

\subsection{Jet Opening Angles and Energetics \label{sec:energ}}
We measure jet opening angles using the methodology of \markcite{burrows07}{Burrows} \& {Racusin} (2007),
originally derived from \markcite{sari99}{Sari} {et~al.} (1999) and \markcite{frail01}{Frail} {et~al.} (2001) where the
opening angle of a uniform jet is defined as:
\begin{equation}\label{eq:thetaj}
  \theta_j=0.057\, \xi\, t^{3/8}_j
\end{equation}
\begin{equation}\label{eq:xi}
  \xi \equiv \left(\frac{3.5}{1+z}\right)^{3/8}
  \left(\frac{\eta_{\gamma}}{0.1}\right)^{1/8}
  \left(\frac{n}{E_{\gamma,iso,53}}\right)^{1/8}
\end{equation}
where $\theta_j$ is the inferred jet half-opening angle, $t_j$ is the jet break
time in days, $z$ is the redshift, $\eta_{\gamma}$ is the assumed radiative
efficiency, $n$ is the ambient number density in $cm^{-3}$, and
$E_{\gamma,iso,53}$ is the rest frame isotropic equivalent energy radiated in gamma rays
between 1 keV and 10 MeV in units of $10^{53}$ ergs.  We assume
$n\sim 1\, cm^{-3}$ in all cases.  The dependence on $n$ is also only
$1/8$, and therefore has only a small effect on $\theta_j$.

Equations \ref{eq:thetaj} and \ref{eq:xi} are only valid for a constant
density (ISM) medium.  For the sake of comparison with pre-\swift values,
and because we find no afterglows that are solely consistent with Wind
medium models, we use only these ISM relations for determining the jet
opening angle.

\markcite{zhang07b}{Zhang} {et~al.} (2007a) measured GRB efficiencies from the X-ray afterglow kinetic
energies and found that GRBs have a distribution of $\eta_{\gamma}$, with 
most showing $\eta_{\gamma}<0.1$.  Therefore, we choose to use a universal
value of $\eta_{\gamma}=0.1$ and make comparisons with the values of
$E_{\gamma}$ listed in the literature \markcite{bloom03,frail01}({Bloom} {et~al.} 2003; {Frail} {et~al.} 2001) for
pre-\swift bursts.  The dependence on 
$\eta_{\gamma}$ is to the $1/8$ power, therefore it has only a weak effect
on the $\theta_j$ estimates. 

\begin{figure}
  \begin{center}
    \includegraphics[scale=0.35,angle=90]{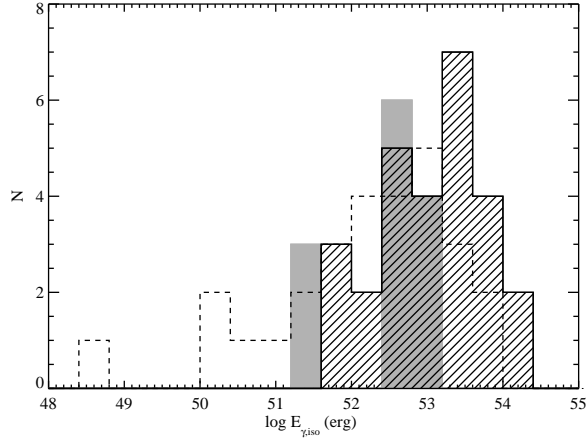}
  \end{center}
  \caption{Distribution of our estimated $E_{\gamma,iso}$s for GRBs
    with measured redshifts in the Prominent
    jet break sample (grey solid), and the Hidden and Possible jet break
    samples (dashed lines),
    compared with the pre-\swift
    measurements (filled hatched histogram) from
    \markcite{bloom03}{Bloom} {et~al.} (2003) with measured redshifts. Note that the 9 SHBs for which we
    were able to estimate $E_{\gamma,iso}$ dominate the low energy end of
    this distribution.
    \label{fig:eiso}}
\end{figure}

Unfortunately $E_{\gamma,iso}$ is often not a well constrained
quantity.  To properly 
measure the bolometric fluence (approximated as the fluence between 1 keV and
10 MeV), coverage to harder energies beyond the hard X-ray band of the BAT
($15-350$ keV)
is needed.  In a handful of cases, simultaneous high energy spectral information is
available in the literature from Konus-Wind or Suzaku that can 
properly characterize the spectra.  We describe our calculations of
$E_{\gamma,iso}$ in Appendix A.
Many assumptions whose error contributions to
$E_{\gamma,iso}$ are unknown go into these calculations.  Therefore
these determinations of $E_{\gamma,iso}$ are to be taken with 
caution.  The dependence on $E_{\gamma,iso}$ in the jet opening angle 
calculations (Eqs. \ref{eq:thetaj} and \ref{eq:xi}) is only to the $1/8$ power,
consequently having minimal impact on 
that quantity, but substantial impact on the estimate of the total collimated
energy output ($E_{\gamma}$).
The distribution of $E_{\gamma,iso}$ for our sample is shown in Figure
\ref{fig:eiso} with the pre-\swift values \markcite{bloom03}({Bloom} {et~al.} 2003) for
comparison. The
\swift $E_{\gamma,iso}$ distribution peaks at and extends to lower energies than
that of the pre-\swift era.  This is probably an effect of the
lower thresholds and softer energy response of the BAT. 

\begin{figure}
  \begin{center}
    \includegraphics[scale=0.35,angle=90]{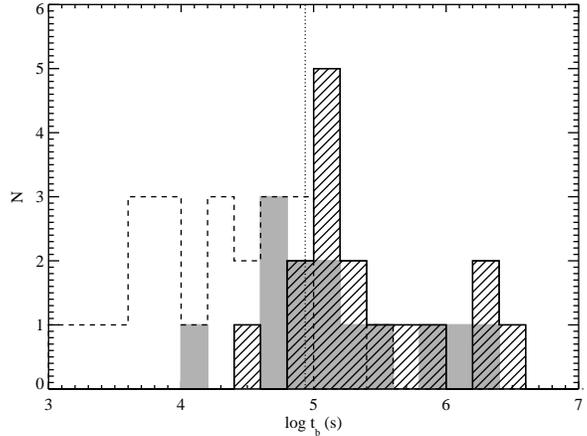}
  \end{center}
  \caption{Distribution of our estimated break times for GRBs with
    measured redshifts in the Prominent
    jet break sample (grey solid), and the Hidden and Possible jet break 
    samples (dashed lines),
    compared with
    the pre-\swift measurements (filled hatched histogram) from
    \markcite{bloom03}{Bloom} {et~al.} (2003). \label{fig:tbreak}}
\end{figure}

\begin{figure}
  \begin{center}
    \includegraphics[scale=0.35,angle=90]{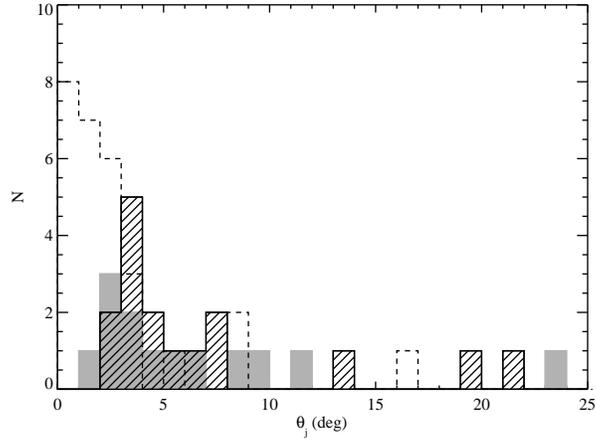}
  \end{center}
  \caption{Distribution of our estimated jet half-opening angles ($\theta_j$)
    for the GRBs with measured redshifts in the Prominent jet break
    sample (grey solid), and the Hidden and Possible 
    jet break samples (dashed lines), compared with the pre-\swift measurements
    (filled hatched histogram) from \markcite{bloom03}{Bloom} {et~al.} (2003). Note that we recalculate
    $\theta_j$ for the pre-\swift sample so that the formalism and density
    estimates are comparable. \label{fig:thetaj}}  
\end{figure}

\begin{figure}
  \begin{center}
    \includegraphics[scale=0.35,angle=90]{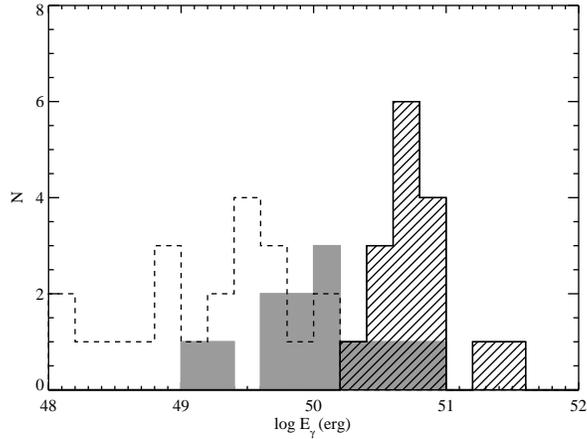}
  \end{center}
  \caption{Distribution of our estimated collimation corrected $\gamma$-ray
    energy ($E_{\gamma}$) for GRBs with measured redshifts in the
    Prominent jet break sample (grey solid), and 
    the Hidden and Possible jet break samples (dashed lines), compared
    with the pre-\swift
    measurements (filled hatched 
    histogram) from \markcite{bloom03}{Bloom} {et~al.} (2003). Note that we recalculate $E_{\gamma}$ for
    the pre-\swift sample using the values of $\theta_j$ shown in Figure
    \ref{fig:thetaj}. 
    \label{fig:egam}}  
\end{figure}

We plot our break times in Figure \ref{fig:tbreak}.
For each burst, using the measured $t_{b}$ and redshift, and estimated 
$E_{\gamma,iso}$, we can determine the jet half-opening angle from 
Eq. \ref{eq:thetaj}.  Our estimated values of $\theta_j$ and the
resulting $E_{\gamma}$ (Eq. \ref{eq:egam}) are 
listed in Tables \ref{table:prominent}-\ref{table:possible} and
plotted in Figures 
\ref{fig:thetaj} and \ref{fig:egam}, along with the pre-\swift values.  
The Prominent jet break distribution (with measured redshift) has a mean
(median) $\theta_j=6.5\ (5.4)$ 
degrees, slightly smaller than the pre-\swift measurements.  The other 
categories of jet breaks have even narrower opening angles corresponding to
even earlier potential jet breaks.  These earlier breaks may be
largely due to post-jet-breaks-with-energy-injection or contamination by
light curve breaks that are not jet breaks at all.

Only 85 of our total 230 GRBs have measured redshifts.  
For candidate jet breaks without measured redshifts (indicated by
dashes in Tables \ref{table:prominent}-\ref{table:possible}), we assume
$z=2.3$ or $z=0.4$, and
$E_{\gamma,iso}=3.7\times10^{52}$ ergs to get an estimate (or limit) on 
$\theta_j$ ($\xi\sim 1.2$; Eq. \ref{eq:xi}).
The redshift values of $z=2.3$ or $z=0.4$ that we assumed for GRBs without
measured redshifts is the mean redshift of \swift long and short GRBs,
respectively for our sample.  These values are similar to those discussed in
the literature \markcite{jakobsson06,bagoly06,fiore07}({Jakobsson} {et~al.} 2006b; {Bagoly} {et~al.} 2006; {Fiore} {et~al.} 2007).  However, the
measurements in the literature were based on the first two years to the
\swift mission when the mean redshift was slightly higher.
More recent estimates suggest that the redshift in the third year of the
\swift mission is lower. This
difference in redshift has only a small impact on the derived opening
angles.  We use the median value of the $E_{\gamma,iso}$ distribution
from those long GRBs with measured redshifts to estimate $\theta_j$ and
$E_{\gamma}$ for those GRBs without measured redshifts.

\begin{figure}
  \begin{center}
    \includegraphics[scale=0.8]{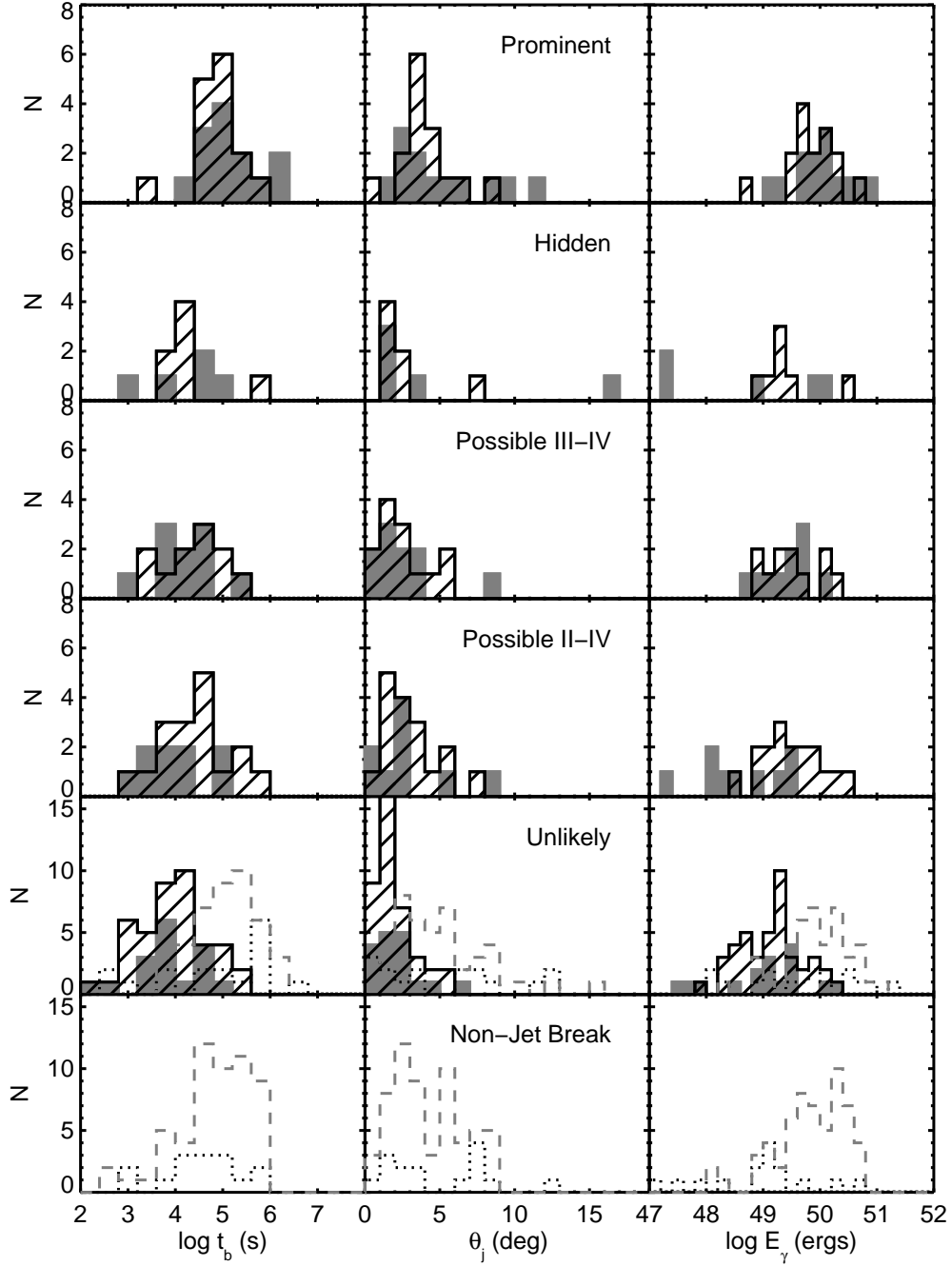}
  \end{center}
  \caption{Distribution of $t_b$ ({\it left}) for all potential jet
  breaks with measured redshifts (dark grey), those without
  measured redshifts (hatched lines).  The jet break lower limit 
  for Unlikely and Non-jet breaks are also shown with (dashed line) and without
  redshifts (dotted line).  Distributions of $\theta_j$ ({\it
    center}), and the collimated energy output, $E_{\gamma}$ ({\it
    right}), are presented for the same samples.  Those GRBs without
  measured redshifts are assumed the average values of $z=2.3$ for long
  bursts or $z=0.4$ for short bursts.
  \label{fig:energetics}}
\end{figure}

For those X-ray afterglows where the last light curve break is consistent with
one of the post-jet break closure relations, we list the jet half-opening angle
($\theta_j$), model requirements, and input parameters in Tables
\ref{table:prominent}-\ref{table:possible}, whether we
have measured or assumed the required relevant parameters.  We show the
distributions of measured $t_b$, $\theta_j$, and $E_{\gamma}$ for all
of those light curves in the 
Prominent, Hidden, Possible, Unlikely, and non-jet break samples in
Figure \ref{fig:energetics}, and compare the distributions 
for those with redshifts and measured $E_{\gamma,iso}$ and those for
which we had to assume average values.  There is no significant
difference between the distributions with and without measured redshifts.

We can now characterize the energy budget of these GRBs.
Using our measurements of $\theta_j$ and $E_{\gamma,iso}$ from the sources
described above, we calculate $E_{\gamma}$, the collimated GRB energies, as:
\begin{equation}\label{eq:egam}
  E_{\gamma}=E_{\gamma,iso}(1-cos\ \theta_j).
\end{equation}
These values and limits are listed in Table
\ref{table:prominent}-\ref{table:nojb}. The distribution of $E_{\gamma}$ is
plotted in the right 
panels of Figure \ref{fig:energetics}. Compared to pre-\swift optical jet break
measurements which tightly cluster around $E_{\gamma}\sim 10^{51}$ ergs
\markcite{bloom03}({Bloom} {et~al.} 2003), our sample is less energetic, with a median value for the
long bursts with estimated $E_{\gamma,iso}$ in Prominent jet break sample of
$\sim 9.8\times 10^{49}$ ergs. 
This measurement is in agreement with that obtained by \markcite{kocevski07}{Kocevski} \& {Butler} (2008). 

\section{Discussion} \label{sec:disc}

There are several different observational categories of potential jet breaks that do not
look like the conventional jet breaks that strictly follow the canonical form
on which previous studies have focused.  These categories
include post-jet break segments with energy injection in segment III where the
normal isotropic models do not fit, apparent segment II-III light curves in
which 
the latter segment requires a post-jet break model suggesting they are actually
segments III-IV, apparent segments II-III that have temporal decays suggestive
of a III-IV or even a II-IV transition, other segments III that require a
post-jet break model and cannot be fit by any of the isotropic models, and
single power-laws that are apparently post-jet break.  Those classifications
for which we are at least somewhat confident are included in the
Prominent, Hidden, and Possible jet break categories.  Those that have
some characteristics suggestive of post-jet break decay, but have temporal
decays are similar to pre-jet break decays are placed in the Unlikely jet
break category.

Perhaps the most unexpected of these categories is the
post-jet-break-with-energy-injection scenario which has been suggested
as an alternative option 
to explain specific GRB afterglows
\markcite{panaitescu06a,oates07,schady07,depasquale08}({Panaitescu} {et~al.} 2006; {Oates} {et~al.} 2007; {Schady} {et~al.} 2007; {de Pasquale} {et~al.} 2008). The idea of energy
injection continuing after the jet break has intriguing implications for
jet break studies.  Energy injection would have the effect of making the
temporal decay shallower than the underlying jet break until the
energy injection ends, which would then manifest as another break in the
light curve. 
This implies that sometimes the break between segments II and III is not caused
by the cessation of energy injection in the isotropic model but rather by
continuation of energy injection through the jet break with the break to
segment IV occurring only after energy injection ceases.  We find 4
examples in the Prominent jet break category and 7 in the Hidden jet
break category which require this scenario to explain our model fits,
in addition to many other segments III that are consistent with
post-jet-break-with-energy-injection models.  This is significantly larger
than the $7.4\pm 1.0$ that we would have expected from our Monte Carlo
simulations.
These specific bursts are indicated in Table
\ref{table:prominent}-\ref{table:hidden} with the 
alternative jet opening angle and jet break time listed for all of those for
which at least one post-jet-break-with-energy-injection model is consistent.

Many single power-law light curves can be explained within the context
of the canonical light curve formalism by observing biases (e.g. late start times,
early stop times).  However, there are several exceptions.
Some simply have large errors in $\alpha$ due to light curves with few
bins. A few well constrained outliers remain, the most 
notable being GRB 061007, whose light curve extends from 80 s to
nearly one million seconds after the trigger, with a continuous,
well constrained, steep, smooth temporal decay.  If we assume that these
afterglows behave like the canonical light curve, these results imply that the
steeper single power-laws 
are post-jet break decays where the preceding light curve segments
were either missed due to a late start or masked by flaring activity.

\begin{figure}
  \begin{center}
    \includegraphics[scale=0.6]{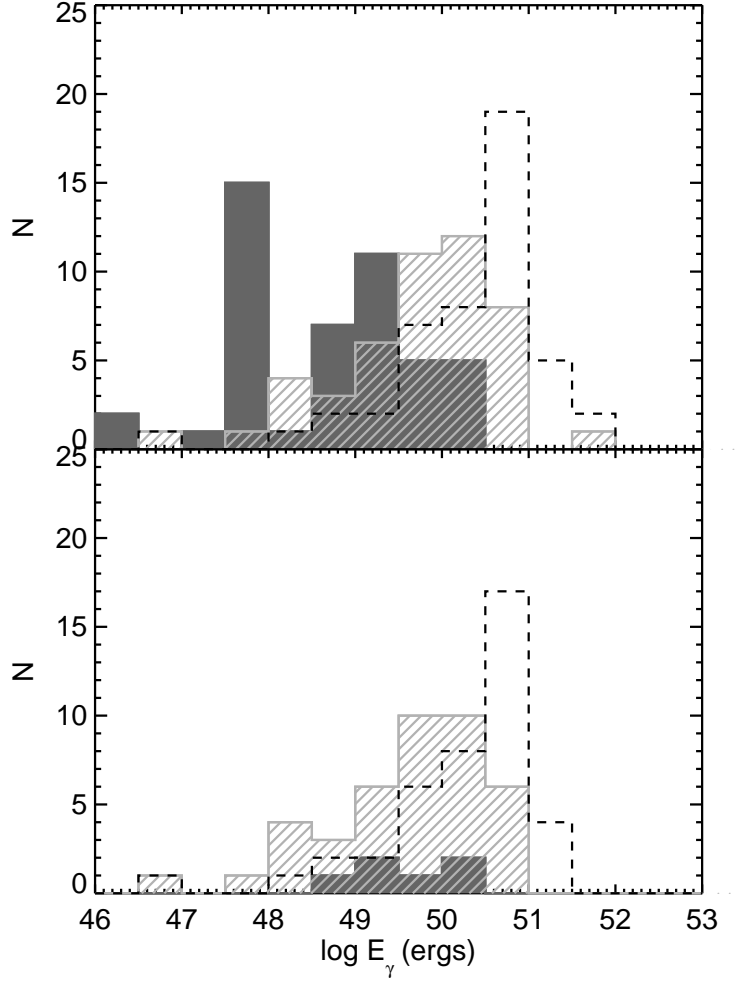}
  \end{center}
   \caption{Distributions of collimated $\gamma$-ray energy output
     limits ({\it top}) for single power-law light curves using jet
     opening angles estimated by using $t_{start}$(solid histogram), 
     $t_{jblim}$ (hatched histogram), $t_{stop}$ (dashed line). 
     The bottom panel also shows the distributions of $E_{\gamma}$
     assuming jet break after $t_{jblim}$ or $t_{stop}$ 
     for post-jet break candidates from Figure \ref{fig:singlepl} and assumes
     pre-jet break limit 
     for remaining single power-law light curves.
     When no redshifts are available, we assume $E_{\gamma,iso}=10^{53}$ ergs and
     $z=2.3$ for long bursts or $z=0.4$ for short bursts.
     \label{fig:singlepl_egam}} 
\end{figure}

In order to gauge the likelihood of these exceptions being pre- or post-jet
break, we calculate the expected collimated energy outputs limits
($E_{\gamma}$) in the method described in \S\ref{sec:energ} for these bursts if
the jet breaks were prior to the start of the observations or after the end of
the observations (Figure \ref{fig:singlepl_egam}).
Those bursts for which the observations extend beyond the time frame for which
we would expect a jet break based on the behavior of the Prominent jet break
category require enormous 
collimated energy outputs ($\gg 10^{51}$ ergs), suggesting that the jet break was
prior to the observation start.  These early jet breaks are difficult 
to explain in terms of the canonical form.  There are 10 of these single
power-law afterglows that persist through the entire expected jet break time
window, 3 of which have very well sampled light curves
(GRB 050716, GRB 061007, GRB 061126).  The 2 other
than GRB 061007 have relatively shallow decays ($\alpha \sim 1$), suggesting
they are pre-jet break.  \markcite{perley08}{Perley} {et~al.} (2008) and \markcite{gomboc08}{Gomboc} {et~al.} (2008)
demonstrate inconsistencies between the optical and X-ray properties
of GRB 061126 suggesting an additional component in needed to explain
the X-rays which is outside of the standard model.  
These bursts remain enigmatic and are difficult to
understood in the context of the majority of \swift burst.  They may have
exceptionally late jet breaks, or perhaps do not break at all, implying a
large jet opening angle or perhaps isotropic outflow.

Another variation on the unusual non-canonical light curve categories are
those with segments I-II where segment II has a slope of $\sim 1$ and is
consistent with the normal spherical decay closure relations as well as the
normal decay with energy injection relations.  There are 19 light curves in
our sample that fit these criteria.  This deviation from the canonical
behavior implies either that these GRBs did not experience the energy injection
phase, or that segment I was misidentified due to flaring behavior. 
These objects may be similar to the shallow single power-law light curves
except that their steep decay was observed first.  In fact, 5 of these light
curves fit our completeness criteria like those exceptional single power-law
cases, implying that they are unusually long lived for segments II or III.
Perhaps the lack of energy injection or cause of this phase differs somehow
from that of the canonical afterglows. It is interesting to note that these
X-ray light curves (after segment I) would have been considered normal in
the pre-\swift era. 

We have unearthed many additional jet break candidates in the data, but the
fundamental question remains: why do more afterglows not have obvious jet
breaks?
The most straightforward and plausible explanation for the lack of conventional
jet breaks in XRT light curves is simply that the observations end before the
jet breaks occur.  
The more fundamental question is what makes these afterglows for which we do
not observe jet breaks different from those for which we do observe jet breaks?

We have some cases of jet breaks at very late times ($>10^6$ s).  At
these times, the fluxes are low, and uncertainties on the data points
are large, making it difficult to detect jet breaks.  In fact, many
light curves end before this time frame, in which case we could be
missing the jet breaks completely.  \markcite{curran08}{Curran} {et~al.} (2008) simulated
GRB afterglow light curves based on real XRT data and showed that
hidden jet breaks could be present in even well sampled XRT light
curves.  We evaluate the probability that such a bias exists within
our data set by doing a similar exercise, calculating the last time at which
a jet break could occur without being detectable.  This time is determined
by forcing an additional break into the light curve with a slope equal to
$\alpha_{f}+1$, where $\alpha_f$ is the measured slope of the last
light curve segment.  We then find the earliest break time that
increases the overall $\chi^2$ by 2.7 (corresponding to $90\%$
confidence for one parameter of interest).  We refer to this time as
the jet break lower limit ($t_{jblim}$).  We excluded light curves
with segment IV from this analysis, and find that an additional $8\%$ of the
light curves could not be fit in this way; in these cases the jet
break limit is the time of the last detection.

\begin{figure}
  \begin{center}
    \includegraphics[scale=0.6]{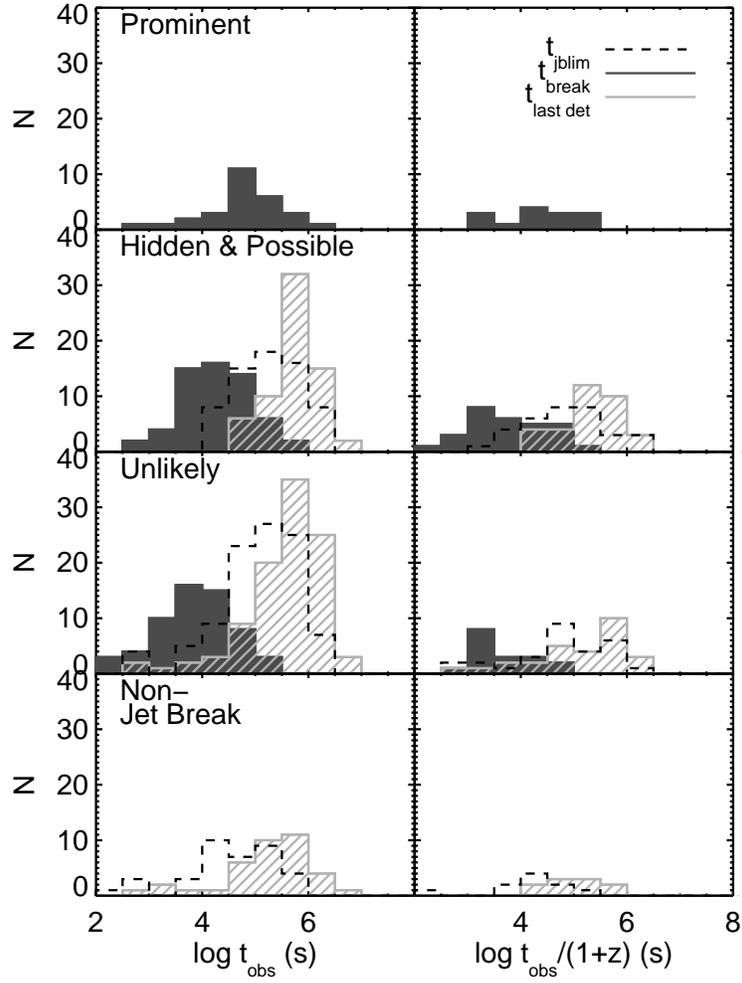}
  \end{center}
  \caption{Distributions of times of the last detection, jet break
    lower limit, and time of potential jet breaks for the
    different categories of jet break in both the observed frame
    ({\it left}) and rest frame ({\it right}).
    \label{fig:tlastdet_dist}}
\end{figure}

The distributions of times of last detections, jet break lower limits,
and times of potential jet breaks in both the observed frame
and rest-frame are shown in Figure \ref{fig:tlastdet_dist}.  The
distributions of jet break times for the 
Prominent, Hidden, and Possible jet break categories overlap, with the latter
sample peaking at an earlier time.  This might be due to contamination in the
Possible jet break sample by breaks that are not jet breaks.  These other
breaks may be from the segment II-III transition, or the previously mentioned 
jet-breaks-with-energy-injection (between segments II and III) which tend to
occur earlier than other (segment III-IV) jet breaks.  The important thing
to extract from Figure \ref{fig:tlastdet_dist} is that the majority of the
bursts in the Unlikely jet break and non-jet break categories have last
detection and jet break lower limits consistent with the range of 
Prominent jet break times.  It is probable that these light curves had jet breaks
after XRT observations ended or that were buried within the noise of the
late time data.  There are a few exceptions to this, particularly in 
those single power-law light curves mentioned above that would have had to
have their jet breaks very early (few $\times 100$ s) and did not break even
beyond the expected jet break times.  Perhaps these few bursts are fundamentally
different from those with jet breaks.

We also included short hard GRBs (SHBs) in our study of jet breaks, treating
them the same as the long bursts except in the calculation of
$E_{\gamma,iso}$.  Only the brightest X-ray afterglows of SHBs were
included in our study, which may bias the understanding of this group in
terms of the faint and quickly fading subsample that did not
meet our minimum requirements.  The SHBs
were placed into similar jet break subsamples as the long GRBs (i.e. 2
Prominent, 4 Possible, 3 Unlikely, and 4 non-jet breaks).  The SHBs have on
average smaller values of $E_{\gamma,iso}$ and $E_{\gamma}$ than the long bursts.
Those SHBs that show jet breaks and similarity to the canonical model perhaps have
some fundamental differences in their environments or physical mechanisms
from those that simply fade quickly \markcite{troja08,sakamoto09}({Troja} {et~al.} 2008; {Sakamoto} \& {Gehrels} 2009).

Comparing this work to other recent studies of jet breaks in X-ray afterglows
\markcite{burrows07,willingale07,panaitescu07,kocevski07,liang08,evans09}({Burrows} \& {Racusin} 2007; {Willingale} {et~al.} 2007; {Panaitescu} 2007; {Kocevski} \& {Butler} 2008; {Liang} {et~al.} 2008; {Evans} {et~al.} 2008), we find
significant overlap with our jet break candidates.  Most differences can be
attributed to differing interpretation of light curve fitting and flaring and
generally more limited jet break definitions in those other studies.
Each study used independent jet break criteria and there were several different
independent data analysis pipelines.  Our work builds upon these other
studies with our systematic analysis of the interaction between all of the
light curve regions for each burst, broad closure relation model usage, careful
searches for jet breaks buried in the data, and characterization of the
energetics and limits.

\section{Conclusions} \label{sec:conc}

Pre-\swift expectations for GRB X-ray afterglows have been
substantially revised with 
the great wealth of XRT observations. While we try to categorize and
classify their properties, there is still a wide range of unexplained diversity
in GRB afterglow properties.  Within the limits of theoretical
expectations and observational biases, we have attempted to survey the
properties of the X-ray afterglows.

In agreement with some previous studies
(\markcite{burrows07}{Burrows} \& {Racusin} 2007, \markcite{liang08}{Liang} {et~al.} 2008, \markcite{willingale07}{Willingale} {et~al.} 2007, \markcite{kocevski07}{Kocevski} \& {Butler} 2008,
\markcite{evans09}{Evans} {et~al.} 2008), we find only a small fraction ($\sim 12\%$) of our total
sample has a late-time break that is clearly a jet break justified by the
closure relations.  We find an additional 
$\sim 30\%$ with observational biases that make segments IV
non-distinct but with a strong case for post-jet 
break temporal and spectral properties.  Some of the bursts in our sample
remain ambiguous in the jet break designation.  Despite not being able to make
absolute claims about these specific bursts, we demonstrate that there are jet
breaks hidden within the data and observational biases.  This suggests that there
are $\sim 20\%$ more jet breaks in the XRT afterglows than previous studies have 
revealed and at least $40\%$ of the missing jet breaks can be attributed to
observational biases.  Some of our light curves that require energy
injection to continue post-jet break may have been previously
misidentified as the end of the energy injection phase.

\markcite{evans09}{Evans} {et~al.} (2008) also explores the canonical X-ray afterglow form using the XRT
sample, where they find that less than half of all light curves behave
canonically, and one quarter are ``oddballs''.  Many of these ``oddballs'' can
be explained by the scenarios we use to describe the ambiguous cases
discussed in this paper.  While their approach 
is somewhat different, their conclusions are similar to ours.

Our study requires post-jet break energy injection to explain 4 cases
of the Prominent jet break sample and 11 others in the Hidden jet break
category.  This modification to the canonical X-ray afterglow form alters
expectations from simply studying the light curve alone, and adds to the
theory needed to explain the diversity of observed properties.

These explanations do not solve all of the remaining problems related to jet
breaks.  Several afterglow light curves, particularly the ones who can
only be fit by a 
single power-law, persist with a constant slope prior to and beyond the times
for which we would expect a jet break.  These bursts require
either an exceptionally early jet break (sometimes before $100$ sec) or an
exceptionally late jet break, requiring a large jet opening angle and an
enormous ($\gg 10^{51}$ erg) collimation corrected energy output.

\swift GRBs are on average at higher redshifts, 
smaller jet opening angles, lower isotropic equivalent energies, and lower
collimated $\gamma$-ray energies compared to GRBs observed prior to {\it Swift}.
Some of these effects can be attributed to the lower energy coverage and
superior sensitivity of the \swift-BAT.  However, the consequences of these
observational biases towards selecting different sorts of bursts is unexpected.

One of the fundamental predictions of the jet break models used in this work is
achromatic behavior in a single spectral component.  The jet break
should be a purely geometrical effect and should 
therefore not be limited to the X-ray afterglows.  Other components of the
afterglow geometry may be more closely tied to emission segments and mechanisms
making direct afterglow comparison difficult.  Modeling of the complete
spectral energy distribution would be necessary to understand how the spectral
breaks might influence the chromatic light curve behavior.  This would be
further complicated by uncertainties in the optical extinction and X-ray
absorption.  These detailed spectral studies, which are beyond the
scope of this work, 
would provide additional information if simultaneous optical,
infrared, and radio observations were available to narrow down the
closure relations models and better constrain the physical models.
\markcite{liang08}{Liang} {et~al.} (2008) have already made some  progress on exploring
multi-wavelength approach to this problem, specifically evaluating
cases of chromatic versus achromatic breaks.

\acknowledgments
J.L.R., D.N.B., and A.F. gratefully acknowledge support for this work from
NASA contract NAS5-00136.  We acknowledge the use of public data from the
Swift data archive. B.Z., B.B.Z, and E.-W.L. gratefully acknowledge support
from NASA contracts NNG05GB67G, NNX08AN24G, NNX08AE57A, and a President's
Infrastructure Award from UNLV. E.-W.L. also acknowledges support from a
National Natural Science Foundation of China grant No. 10873002, National
Basic Research Program ("973" Program) of China (Grant 2009CB824800), and
the research foundation of Guangxi University. This work made use of data
supplied by the UK Swift Science Data Centre at the University of
Leicester.  We thank D. Willingale for his helpful comments.

\bibliography{}


\clearpage

\begin{deluxetable}{lcccccc} 
\tablecolumns{8} 
\tablewidth{0pc}
\tablecaption{Closure Relations \label{table:crs}} 
\tabletypesize{\footnotesize}
\tablehead{ 
\colhead{} & \colhead{} & \colhead{} & a & b & \colhead{} & c \\
\colhead{} & \colhead{} & \colhead{} &  \multicolumn{2}{c}{No Energy Injection} & &
\colhead{Energy Injection} \\
\cline{4-5} \cline{7-7}
\colhead{} & \colhead{} &  \colhead{$\beta$} & \colhead{$\alpha(\beta)$} &
\colhead{$\alpha(\beta)$} & & \colhead{$\alpha(\beta)$} \\
\colhead{} & \colhead{} & \colhead{} & \colhead{$(p>2)$} & \colhead{$(1<p<2)$} & & 
\colhead{$(p>2)$}}
\startdata
\\[-0.7cm]
\multicolumn{7}{c}{ISM, Slow Cooling} \\[0cm]
\cline{1-7}
1 & $\nu_m<\nu<\nu_c$ & $\frac{p-1}{2}$ & $\alpha=\frac{3\beta}{2}$  &
$\alpha=\frac{3(2\beta+3)}{16}$ & & $\alpha=(q-1)+\frac{(2+q)\beta}{2}$ \\
2 & $\nu>\nu_c$ & $\frac{p}{2}$ & $\alpha=\frac{3\beta-1}{2}$ &  $\alpha=\frac{3\beta+5}{8}$
& & $\alpha=\frac{q-2}{2}+\frac{(2+q)\beta}{2}$ \\
\cline{1-7}
\multicolumn{7}{c}{ISM, Fast Cooling} \\[0cm]
\cline{1-7}
3 & $\nu_c<\nu<\nu_m$ & $\frac{1}{2}$ & {\tiny \textsl{(3a)}} $\alpha=\frac{\beta}{2}$ &
{\tiny \textsl{(3b)}} $\alpha=\frac{\beta}{2}$ & & {\tiny \textsl{(3c)}} $\alpha=(q-1)+\frac{(2-q)\beta}{2}$ \\
4 & $\nu>\nu_m$ & $\frac{p}{2}$ & {\tiny \textsl{(4a)}} $\alpha=\frac{3\beta-1}{2}$ &
{\tiny \textsl{(4b)}} $\alpha=\frac{3\beta+5}{8}$ & & {\tiny \textsl{(4c)}} $\alpha=\frac{q-2}{2}+\frac{(2+q)\beta}{2}$ \\
\cline{1-7}
\multicolumn{7}{c}{Wind, Slow Cooling} \\
\enddata
\end{deluxetable}


\begin{deluxetable}{lllccrccrrl}
\tabletypesize{\scriptsize}
\tablecaption{Prominent Jet Breaks \label{table:prominent}}
\tablewidth{0pt}
\tablehead{
  \colhead{GRB} & \colhead{Segments} & \colhead{Time} & \colhead{$\alpha$} & \colhead{$\beta$} & \colhead{t} & \colhead{z} & \colhead{log $E_{\gamma,iso}$} & \colhead{$\theta_j$} &   \colhead{log $E_{\gamma}$} & \colhead{Requirements\tablenotemark{\dag}} \\
 \colhead{} &  \colhead{} &  \colhead{} &  \colhead{} &  \colhead{} &  \colhead{(ks)} &  \colhead{} &  \colhead{(ergs)} &  \colhead{(deg)} &  \colhead{(ergs)} &  \colhead{}} 
\startdata
050315  & I-II-III-IV & $t_{b}$ & $1.86^{+0.58}_{-0.36}$ & $1.30^{+0.35}_{-0.29}$ & $240.6^{+69.3}_{-76.3}$ & $1.95$\tablenotemark{a} & $52.7$ & $5.6$ & 50.3 &  Uni $\nu_2$ $p_2$ \\
           &  & $t_{b,EI}$ & $0.69^{+0.05}_{-0.05}$ & $0.96^{+0.09}_{-0.09}$ & $7.3^{+1.4}_{-1.1}$ &  &  & $1.5$ & 49.2 &  \textbf{EI} $\nu_2$ $p_2$ \\
050319  & I-II-III-IV & $t_{b}$ & $1.58^{+0.44}_{-0.26}$ & $1.37^{+0.57}_{-0.47}$ & $55.3^{+19.1}_{-26.0}$ & $3.24$\tablenotemark{b} & $52.6$ & $2.8$ & 49.7 &  Uni $\nu_2$ \\
           &  & $t_{b,EI}$ & $0.67^{+0.15}_{-0.14}$ & $0.95^{+0.15}_{-0.14}$ & $4.0^{+7.6}_{-2.2}$ &  &  & $1.1$ & 48.9 &  \\
050505  & II-III-IV & $t_{b}$ & $1.81^{+0.30}_{-0.15}$ & $1.13^{+0.17}_{-0.16}$ & $45.9^{+25.4}_{-11.0}$ & $4.28$\tablenotemark{c} & $53.2$ & $2.1$ & 50.0 &  $\nu_2$ \\
           &  & $t_{b,EI}$ & $1.12^{+0.10}_{-0.09}$ & $1.07^{+0.13}_{-0.12}$ & $6.5^{+2.4}_{-1.1}$ &  &  & $1.0$ & 49.4 &  \\
050713B & I-II-III-IV & $t_{b}$ & $2.10^{+1.91}_{-0.71}$ & $1.13^{+7.87}_{-2.13}$ & $381.2^{+314.9}_{-114.3}$ &  --  &  --  & $6.5  \xi$ &  --  &  Uni \\
           &  & $t_{b,EI}$ & $1.04^{+0.13}_{-0.14}$ & $0.92^{+0.20}_{-0.23}$ & $32.9^{+12.2}_{-20.5}$ &  --  &  --  & $2.6  \xi$ &  --  &  \\
\enddata
\tablenotetext{\dag}{Abbreviations for Uniform jet (Uni), Spreading jet (Sp), non-spreading jet (NSp), $1<p<2$ ($p_{12}$), $p>2$ ($p_2$), $\nu<\nu_c$ ($\nu_1$), $\nu>\nu_c$ ($\nu_2$)} 
\tablecomments{Prominent jet breaks are those with a distinct segment IV which are consistent with the post-jet break closure relations.  Alternative times for jet angle limits depend on models fit.  Energy injection (EI) breaks are listed only if at least one jet-break-with-energy-injection-relation is consistent.  Requirements listed are for those properties of families of closure relations that are required by consistent models.  $\xi$ is defined in Equation \ref{eq:xi}.  All errors are $2\sigma$ confidence.}
\tablerefs{ $^{a}$\markcite{berger05}{Berger} {et~al.} (2005b), $^{b}$\markcite{jakobsson06b}{Jakobsson} {et~al.} (2006a), $^{c}$\markcite{berger06}{Berger} {et~al.} (2006b), $^{d}$\markcite{jakobsson06}{Jakobsson} {et~al.} (2006b), $^{e}$\markcite{prochaska07}{Prochaska} {et~al.} (2007a), $^{f}$\markcite{GCN4186}{Soderberg}, {Berger}, \& {Ofek} (2005), $^{g}$\markcite{GCN4221}{Quimby} {et~al.} (2005), $^{h}$\markcite{soderberg06}{Soderberg} {et~al.} (2006a), $^{i}$\markcite{ferrero09}{Ferrero} {et~al.} (2008), $^{j}$\markcite{galyam06}{Soderberg} {et~al.} (2006b), $^{k}$\markcite{jakobsson06b}{Jakobsson} {et~al.} (2006a), $^{l}$\markcite{GCN5373}{Th\"one} {et~al.} (2006b), $^{m}$\markcite{GCN6663}{Th\"one}, {Perley}, \&  {Bloom} (2007c), $^{n}$\markcite{jaunsen08}{Jaunsen} {et~al.} (2008)}
\end{deluxetable}

\begin{deluxetable}{lllccrccrrl}
\tabletypesize{\scriptsize}
\tablecaption{Hidden Jet Breaks \label{table:hidden}}
\tablewidth{0pt}
\tablehead{
  \colhead{GRB} & \colhead{Segments} & \colhead{Time} & \colhead{$\alpha$} & \colhead{$\beta$} & \colhead{t} & \colhead{z} & \colhead{log $E_{\gamma,iso}$} & \colhead{$\theta_j$} &   \colhead{log $E_{\gamma}$} & \colhead{Requirements\tablenotemark{\dag}} \\
 \colhead{} &  \colhead{} &  \colhead{} &  \colhead{} &  \colhead{} &  \colhead{(ks)} &  \colhead{} &  \colhead{(ergs)} &  \colhead{(deg)} &  \colhead{(ergs)} &  \colhead{}} 
\startdata
050416A & I-II-III & $t_{b}$ & $0.90^{+0.04}_{-0.03}$ & $0.40^{+0.10}_{-0.10}$ & $1.1^{+1.0}_{-0.4}$ & $0.65$\tablenotemark{a} & $50.8$ & $1.6$ & 47.4 &  \textbf{EI} $\nu_1$ $p_2$ \\
050802  & II-IV & $t_{b}$ & $1.52^{+0.13}_{-0.06}$ & $0.85^{+0.07}_{-0.13}$ & $6.1^{+3.1}_{-1.2}$ & $1.71$\tablenotemark{b} & $52.4$ & $1.6$ & 49.0 &  \textbf{EI} $\nu_1$ $p_2$ \\
050815  & II-IV & $t_{b}$ & $2.21^{+0.51}_{-0.32}$ & $0.39^{+0.65}_{-0.47}$ & $4.3^{+0.7}_{-0.8}$ &  --  &  --  & $1.2  \xi$ &  --  &  $\nu_1$ \\
           &  & $t_{b,EI}$ & $0.19^{+0.14}_{-0.17}$ & $1.15^{+0.92}_{-0.51}$ & $<0.09$ &  --  &  --  & $0.3  \xi$ &  --  &  \\
050822  & I-II-III & $t_{b}$ & $1.04^{+0.06}_{-0.06}$ & $1.26^{+0.20}_{-0.18}$ & $18.1^{+4.9}_{-4.2}$ &  --  &  --  & $2.1  \xi$ &  --  &  \textbf{EI} $p_2$ \\
051111  & I-II-IV & $t_{b}$ & $4.78^{+1.93}_{-1.49}$ & $1.22^{+0.40}_{-0.37}$ & $42.2^{+13.4}_{-8.7}$ & $1.55$\tablenotemark{c} & $52.7$ & $3.0$ & 49.9 &  Sp $\nu_1$ $p_2$ \\
060210  & II-III & $t_{b}$ & $1.30^{+0.09}_{-0.07}$ & $1.07^{+0.10}_{-0.10}$ & $25.8^{+9.8}_{-6.9}$ & $3.91$\tablenotemark{d} & $53.6$ & $1.5$ & 50.1 &  \textbf{EI} $p_2$ \\
060218  & I-II-IV & $t_{b}$ & $1.30^{+0.34}_{-0.21}$ & $4.86^{+2.59}_{-1.00}$ & $71.2^{+132.1}_{-59.4}$ & $0.03$\tablenotemark{e} & $48.7$ & $16.4$ & 47.3 &  NSp \textbf{EI} $\nu_2$ $p_2$ \\
\enddata
\tablenotetext{\dag}{Abbreviations for Uniform jet (Uni), Spreading jet (Sp), non-spreading jet (NSp), $1<p<2$ ($p_{12}$), $p>2$ ($p_2$), $\nu<\nu_c$ ($\nu_1$), $\nu>\nu_c$ ($\nu_2$)} 
\tablecomments{Hidden jet breaks require jet break closure relations in the last light curve segment.  $t_{start}$ parameters are listed for those afterglows that are consistent with a jet-break-with-energy-injection during the first segment, suggesting that the jet break could have occurred prior to the start of the observations. Other notes are the same as for Prominent Jet Breaks table.}
\tablerefs{ $^{a}$\markcite{soderberg07}{Soderberg} {et~al.} (2007), $^{b}$\markcite{GCN3749}{Fynbo} {et~al.} (2005), $^{c}$\markcite{prochaska06}{Prochaska}, {Chen}, \&  {Bloom} (2006b), $^{d}$\markcite{GCN4729}{Cucchiara}, {Fox}, \&  {Berger} (2006a), $^{e}$\markcite{mirabal06}{Mirabal} {et~al.} (2006)}
\end{deluxetable}

\begin{deluxetable}{lllccrccrrl}
\tabletypesize{\scriptsize}
\tablecaption{Possible Jet Breaks \label{table:possible}}
\tablewidth{0pt}
\tablehead{
  \colhead{GRB} & \colhead{Segments} & \colhead{Time} & \colhead{$\alpha$} & \colhead{$\beta$} & \colhead{t} & \colhead{z} & \colhead{log $E_{\gamma,iso}$} & \colhead{$\theta_j$} &   \colhead{log $E_{\gamma}$} & \colhead{Requirements\tablenotemark{\dag}} \\
 \colhead{} &  \colhead{} &  \colhead{} &  \colhead{} &  \colhead{} &  \colhead{(ks)} &  \colhead{} &  \colhead{(ergs)} &  \colhead{(deg)} &  \colhead{(ergs)} &  \colhead{}} 
\startdata
050219A & I-II-III & $t_{b}$ & $1.29^{+0.92}_{-0.28}$ & $0.91^{+0.26}_{-0.24}$ & $32.3^{+114.2}_{-26.0}$ &  --  &  --  & $2.6  \xi$ &  --  &  \\
           &  & $t_{lastdet}$ & $1.29^{+0.92}_{-0.28}$ & $0.91^{+0.26}_{-0.24}$ & $>3091.94$ &  --  &  --  & $>14.3  \xi$ &  --  &  \\
050318  & III-IV & $t_{b}$ & $2.06^{+0.58}_{-0.35}$ & $1.02^{+0.22}_{-0.20}$ & $20.7^{+10.2}_{-9.6}$ & $1.44$\tablenotemark{a} & $51.9$ & $2.9$ & 49.1 &  \\
           &  & $t_{lastdet}$ & $2.06^{+0.58}_{-0.35}$ & $1.02^{+0.22}_{-0.20}$ & $>57.83$ &  &  & $>4.3$ & $>49.4$ &  \\
           &  & $t_{start}$ & $1.35^{+0.12}_{-0.16}$ & $1.13^{+0.18}_{-0.16}$ & $<3.28$ &  &  & $<1.5$ & $<48.5$ &  \\
050326  & III-IV & $t_{b}$ & $1.73^{+0.30}_{-0.21}$ & $0.88^{+0.42}_{-0.24}$ & $27.0^{+45.3}_{-6.1}$ &  --  &  --  & $2.4  \xi$ &  --  &  \\
           &  & $t_{lastdet}$ & $1.73^{+0.30}_{-0.21}$ & $0.88^{+0.42}_{-0.24}$ & $>531.42$ &  --  &  --  & $>7.4  \xi$ &  --  &  \\
           &  & $t_{start}$ & $1.51^{+0.16}_{-0.61}$ & $1.11^{+0.71}_{-0.59}$ & $<3.26$ &  --  &  --  & $<1.1  \xi$ &  --  &  \\
\enddata
\end{deluxetable}

\begin{deluxetable}{lllccrccrrl}
\tabletypesize{\scriptsize}
\tablecaption{Unlikely Jet Breaks \label{table:unlikely}}
\tablewidth{0pt}
\tablehead{
  \colhead{GRB} & \colhead{Segments} & \colhead{Time} & \colhead{$\alpha$} & \colhead{$\beta$} & \colhead{t} & \colhead{z} & \colhead{log $E_{\gamma,iso}$} & \colhead{$\theta_j$} &   \colhead{log $E_{\gamma}$} & \colhead{Requirements\tablenotemark{\dag}} \\
 \colhead{} &  \colhead{} &  \colhead{} &  \colhead{} &  \colhead{} &  \colhead{(ks)} &  \colhead{} &  \colhead{(ergs)} &  \colhead{(deg)} &  \colhead{(ergs)} &  \colhead{}} 
\startdata
050124  & Single PL & $t_{lastdet}$ & $1.50^{+0.21}_{-0.16}$ & $0.83^{+0.30}_{-0.18}$ & $>4967.45$ &  --  &  --  & $>17.1  \xi$ &  --  &  \\
           &  & $t_{start}$ & $1.50^{+0.21}_{-0.16}$ & $0.83^{+0.30}_{-0.18}$ & $<11.13$ &  --  &  --  & $<1.7  \xi$ &  --  &  \\
050128  & II-III & $t_{b}$ & $1.30^{+0.08}_{-0.07}$ & $0.96^{+0.17}_{-0.16}$ & $2.9^{+3.9}_{-1.0}$ &  --  &  --  & $1.0  \xi$ &  --  &  \\
           &  & $t_{lastdet}$ & $1.30^{+0.08}_{-0.07}$ & $0.96^{+0.17}_{-0.16}$ & $>99.44$ &  --  &  --  & $>4.0  \xi$ &  --  &  \\
           &  & $t_{start}$ & $0.70^{+0.16}_{-0.20}$ & $0.77^{+0.24}_{-0.17}$ & $<0.24$ &  --  &  --  & $<0.4  \xi$ &  --  &  \\
050215B & Single PL & $t_{lastdet}$ & $0.93^{+0.21}_{-0.17}$ & $0.66^{+0.63}_{-0.49}$ & $>3011.04$ &  --  &  --  & $>14.2  \xi$ &  --  &  \\
           &  & $t_{start}$ & $0.93^{+0.21}_{-0.17}$ & $0.66^{+0.63}_{-0.49}$ & $<5.81$ &  --  &  --  & $<1.4  \xi$ &  --  &  \\
050219B & Single PL & $t_{lastdet}$ & $1.21^{+0.05}_{-0.05}$ & $1.11^{+0.19}_{-0.17}$ & $>120.17$ &  --  &  --  & $>4.2  \xi$ &  --  &  \\
\enddata
\end{deluxetable}

\begin{deluxetable}{lllccrccrrl}
\tabletypesize{\scriptsize}
\tablecaption{Non-Jet Breaks \label{table:nojb}}
\tablewidth{0pt}
\tablehead{
  \colhead{GRB} & \colhead{Segments} & \colhead{Time} & \colhead{$\alpha$} & \colhead{$\beta$} & \colhead{t} & \colhead{z} & \colhead{log $E_{\gamma,iso}$} & \colhead{$\theta_j$} &   \colhead{log $E_{\gamma}$} & \colhead{Requirements\tablenotemark{\dag}} \\
 \colhead{} &  \colhead{} &  \colhead{} &  \colhead{} &  \colhead{} &  \colhead{(ks)} &  \colhead{} &  \colhead{(ergs)} &  \colhead{(deg)} &  \colhead{(ergs)} &  \colhead{}} 
\startdata
050126  & I-II & $t_{lastdet}$ & $0.93^{+0.23}_{-0.43}$ & $0.85^{+0.74}_{-0.52}$ & $>70.49$ & $1.29$\tablenotemark{a} & $51.9$ & $>0.7$ & $>47.8$ &  \\
050406  & I-II & $t_{lastdet}$ & $0.61^{+0.24}_{-0.25}$ & $1.32^{+0.83}_{-0.52}$ & $>1000.88$ & $2.44$\tablenotemark{b} & $51.2$ & $>1.3$ & $>47.6$ &  \\
050421  & II-III & $t_{lastdet}$ & $3.42^{+0.28}_{-0.25}$ & $0.63^{+0.58}_{-0.47}$ & $>1.34$ &  --  &  --  & $>0.3  \xi$ &  --  &  none \\
050422  & I-II & $t_{lastdet}$ & $0.83^{+0.24}_{-0.39}$ & $2.03^{+5.26}_{-1.51}$ & $>1324.00$ &  --  &  --  & $>0.5  \xi$ &  --  &  \\
050721  & I-II & $t_{lastdet}$ & $1.17^{+0.09}_{-0.23}$ & $1.04^{+0.23}_{-0.31}$ & $>3385.42$ &  --  &  --  & $>0.6  \xi$ &  --  &  \\
050819  & I-II & $t_{lastdet}$ & $0.40^{+0.19}_{-0.16}$ & $1.46^{+0.82}_{-0.59}$ & $>410.41$ &  --  &  --  & $>0.6  \xi$ &  --  &  \\
050904  & Single PL & $t_{lastdet}$ & $2.07^{+0.11}_{-0.09}$ & $0.19^{+0.06}_{-0.06}$ & $>313.51$ & $6.30$\tablenotemark{c} & $53.7$ & $>3.3$ & $>50.9$ &  \\
050908  & I-II & $t_{lastdet}$ & $1.08^{+0.11}_{-0.11}$ & $1.21^{+0.45}_{-0.43}$ & $>121.71$ & $3.34$\tablenotemark{d} & $52.2$ & $>0.4$ & $>47.6$ &  \\
\enddata
\tablenotetext{\dag}{Abbreviations for Uniform jet (Uni), Spreading jet (Sp), non-spreading jet (NSp), $1<p<2$ ($p_{12}$), $p>2$ ($p_2$), $\nu<\nu_c$ ($\nu_1$), $\nu>\nu_c$ ($\nu_2$)} 
\tablecomments{Non-jet breaks are the remaining GRBs not in the other categories that are with inconsistent with previous criteria, only contain segments I-II, are naked GRBs (0-I), or were ruled out by internal consistency checks. $t_{start}$ and $t_{lastdet}$ times and opening angle limits are only listed if they are feasible within the canonical framework.  Other notes are the same as for Prominent Jet Breaks table.}
\tablerefs{ $^{a}$\markcite{berger05b}{Berger} {et~al.} (2005a), $^{b}$\markcite{schady06} (), $^{c}$\markcite{kawai06}{Kawai} {et~al.} (2006), $^{d}$\markcite{GCN3948}{Fugazza} {et~al.} (2005), $^{e}$\markcite{price07}{Price} {et~al.} (2007), $^{f}$\markcite{GCN5958}{Berger} (2006), $^{g}$\markcite{GCN6010}{Jaunsen} {et~al.} (2007b), $^{h}$\markcite{GCN6322}{Cenko} {et~al.} (2007b), $^{i}$\markcite{GCN6665}{Cucchiara} {et~al.} (2007b), $^{j}$\markcite{davanzo09}{D'Avanzo} {et~al.} (2009)}
\end{deluxetable}

\begin{deluxetable}{lcll}
\tabletypesize{\scriptsize}
\tablecaption{Jet Break Summary\label{table:summary}}
\tablewidth{0pt}
\tablehead{ 
  \colhead{Classification} & \colhead{GRBs} & \colhead{Segment Types} & \colhead{Definition}}
\startdata
Prominent jet break & 28 & I-II-III-IV, II-III-IV & Has distinct segment IV\\
Hidden jet break & 12 & Single power-law, Ambiguous   &
Last segment requires post-jet break closure relation \\
& & II-III/III-IV, I-II-III & \\
Possible jet break & 53 & Ambiguous II-III/III-IV,  &
Resembles III-IV or II-IV transition in $\alpha-\alpha$ space\\
& & I-II-III, Single power-law & \\
Unlikely jet break & 100 & Single power-law, Ambiguous  &
Consistent with both pre- and post-jet break closure \\ 
& & II-III/III-IV, I-II-III & relations, but resembles II-III in
$\alpha-\alpha$ space, or no \\ 
& & &  closure relations but temporally consistent\\
Non-jet break & 37 & I-II, Single power-law, II-III, 0-I & Does not fit any jet break
criteria \\ 

\enddata
\end{deluxetable}

\clearpage
\appendix

\section{Appendix}
To measure $E_{\gamma,iso}$, we must know (or
estimate) the shape of the spectrum and integrate it over the desired energy
range, correcting for the distance and redshift (k-correction) effects using the
method of \markcite{amati02}{Amati} {et~al.} (2002)
\begin{equation}\label{eq:eiso}
  E_{\gamma,iso}=\frac{4\pi D_L^2}{(1+z)}\int_{1~keV/(1+z)}^{10~MeV/(1+z)} E~F(E)~dE
\end{equation}
where $F(E)$ is the functional form of the spectrum, and $D_L$ is the
luminosity distance. 

We assume that the Band function \markcite{band93}({Band} {et~al.} 1993) represents the intrinsic shape
of the spectra for long GRBs even if the BAT data are insufficient to constrain
a fit to this model.  We choose to assume
or infer the parameters of the Band function, and use its functional form
rather than use the fits to the power-law or cutoff power-law due to their
gross over-estimates of the bolometric fluence. 

The Band function takes the form
\begin{equation}\label{eq:band}
  F(E)=
  \begin{cases}
    K^B_{50}\left(\frac{E}{50 ~keV}\right)^{\alpha_B}
    exp\left(-\frac{E}{E_0}\right) & (\alpha_B-\beta_B)E_0 \geq  E \\
    K^B_{50}\left(\frac{(\alpha_B-\beta_B)E_0}{50 ~keV}\right)^{\alpha_B-\beta_B}
    exp(\beta_B-\alpha_B)\left(\frac{E}{50 ~keV}\right)^{\beta_B} & (\alpha_B-\beta_B)E_0 \leq  E \\
  \end{cases}
\end{equation}
where  $K^{B}_{50}$ is the normalization at $50$ keV, and $\alpha_B$ and
$\beta_B$ are the spectral slopes below and above the 
break energy ($E_0$).  When available, we use the measured spectral
slopes, otherwise we assume $\alpha_B=-1$ and $\beta_B=-2.5$.  The relationship
between the spectral peak energy and the Band break energy is
$E_{p}=(2+\alpha_B)E_0$.  

Unfortunately the data are often not sufficient to measure the Band
function parameters because of the limited observed bandpass and
measurement uncertainties.  In some cases, better fits can be obtained
using an exponential cutoff power-law (CPL) model, and we can use these fits
to estimate $E_p$ for input into the Band function.  The CPL takes the form  
\begin{equation}\label{eq:cpl}
  F(E)=K^{CPL}_{50}\left(\frac{E}{50~keV}\right)^{\alpha_{CPL}}
  exp\left(\frac{-E(2+\alpha_{CPL})}{E_p}\right)
\end{equation}
where  $K^{CPL}_{50}$ is the normalization at 50 keV, and $\alpha_{CPL}$ is
the spectral index.  If available in the
literature, we obtain measurements of the peak energy ($E_p$) from either Band
function or exponential cutoff power-law fits to the $\gamma$-ray data
(Table \ref{table:grbpars}).

An adequate CPL fit and therefore an $E_p$ measurement is available for only
$\sim28\%$ of the GRBs in our sample. 
For those without a directly measured $E_p$ from either the Band or CPL
fits, we infer a value using the 
correlation between $E_p$ and the spectral index from fitting a single
power-law ($\alpha_{PL}$) to the BAT data as parametrized by \markcite{zhang07b}{Zhang} {et~al.} (2007a) as
\begin{equation}\label{eq:epeak}
  log~E_p=2.76-3.61~log(-\alpha_{PL})
\end{equation}
where $\alpha_{PL}$ comes from fitting a simple power-law function of the form
\begin{equation}\label{eq:pl}
  F(E)=K^{PL}_{50}\left(\frac{E}{50~keV}\right)^{\alpha_{PL}},
\end{equation}
and $K^{PL}_{50}$ is the normalization at 50 keV.  We use the measurements
of $\alpha_{PL}$ and $\alpha_{CPL}$ from \markcite{sakamoto07}{Sakamoto} {et~al.} (2008).  However, this
method only works for those GRBs whose $E_p$ is inside the BAT energy
band indicated by 
$-2.3<\alpha_{PL}<-1.2$.  We are unable to estimate $E_{\gamma,iso}$ for those GRBs
with $\alpha_{PL}$ outside this range that do not have a measurement of
$E_p$ by another instrument.

Short burst spectra are better characterized by the
exponential cutoff power-law model than the Band function.  They also tend
to have harder photon indices ($\sim 
0.8$) than long bursts.  Therefore we assume the cutoff power-law model to
describe the gamma-ray spectra for short bursts in this sample used to
calculate $E_{\gamma,iso}$ and assume $\alpha_{CPL}=-0.8$ if no measurement
is available.

\renewcommand{\thefootnote}{\fnsymbol{footnote}}
\begin{deluxetable}{lrrrrr}
\tabletypesize{\scriptsize}
\tablecaption{GRBs spectral properties from the literature \label{table:grbpars}}
\tablewidth{0pt}
\tablehead{
  \colhead{GRB} & \colhead{$\alpha_B$} & \colhead{$\beta_B$} &
  \colhead{$\alpha_{CPL}$} & \colhead{$E_p $(keV)} & \colhead{Source}}
\startdata
050318 &  &  &  & $49\pm 7$ & \markcite{still05}{Still} {et~al.} (2005)\\
050406 &  &  &  & $24$ & \markcite{zhang07b}{Zhang} {et~al.} (2007a)\\
050416A &  &  &  & $15.6^{+2.3}_{-2.7}$ & \markcite{sakamoto06}{Sakamoto} {et~al.} (2006)\\
050525A &  &  &  & $82^{+4}_{-3}$ & \markcite{sakamoto07}{Sakamoto} {et~al.} (2008)\\
050603 & $-0.79 \pm 0.06$ & $-2.15 \pm 0.09$ &  & $349\pm 28$ & \markcite{GCN3518}{Golenetskii} {et~al.} (2005a)\\
051109A &  &  &  & $161^{+224}_{-58}$ & \markcite{GCN4238}{Golenetskii} {et~al.} (2005b)\\
051221A\tablenotemark{*} &  &  & $-1.08^{+0.13}_{-0.14}$ & $402^{+93}_{-72}$ & \markcite{GCN4394}{Golenetskii} {et~al.} (2005c)\\
060115 &  &  &  & $63^{+36}_{-11}$ & \markcite{sakamoto07}{Sakamoto} {et~al.} (2008)\\
060124 &  &  &  & $193^{+38}_{-39}$ & \markcite{romano06}{Romano} {et~al.} (2006)\\
060206 &  &  &  & $78^{+38}_{-13}$ & \markcite{sakamoto07}{Sakamoto} {et~al.} (2008)\\
060418 &  &  &  & $230$ & \markcite{GCN4989}{Golenetskii} {et~al.} (2006f)\\
060614\tablenotemark{*} &  &  & $-1.57^{+0.12}_{-0.14}$ & $302^{+214}_{-85}$ & \markcite{GCN5264}{Golenetskii} {et~al.} (2006b)\\
060707 &  &  &  & $63^{+21}_{-10}$ & \markcite{sakamoto07}{Sakamoto} {et~al.} (2008)\\
060814 &  &  &  & $257^{+122}_{-58}$ & \markcite{GCN5460}{Golenetskii} {et~al.} (2006c)\\
060908 &  &  &  & $151^{+184}_{-41}$ & \markcite{sakamoto07}{Sakamoto} {et~al.} (2008)\\
060927 &  &  &  & $72^{+25}_{-11}$ & \markcite{sakamoto07}{Sakamoto} {et~al.} (2008)\\
061007 & $-0.7\pm 0.04$ & $-2.61^{+0.15}_{-0.21}$ &  & $399^{+19}_{-18}$ & \markcite{GCN5722}{Golenetskii} {et~al.} (2006d)\\
061121 & $-0.83^{+0.24}_{-0.19}$ & $-2.00^{+0.18}_{-0.32}$ &  & $455 \pm 115$ & \markcite{GCN5837}{Golenetskii} {et~al.} (2006a)\\
061201\tablenotemark{*} &  &  & $-0.36^{+0.40}_{-0.65}$ & $873^{+458}_{-284}$ & \markcite{GCN5890}{Golenetskii} {et~al.} (2006e)\\
070125 & $-1.1^{+0.10}_{-0.09}$ & $-2.08^{+0.10}_{-0.15}$ & &
$367^{+67}_{-51}$ & \markcite{bellm08}{Bellm} {et~al.} (2008) \\
070508 &  &  &  & $188 \pm 8$ & \markcite{GCN6403}{Golenetskii} {et~al.} (2007d)\\
070714B\tablenotemark{*} &  &  & $-0.86 \pm 0.10$ & $1120^{+780}_{-380}$ & \markcite{GCN6638}{Ohno} {et~al.} (2007)\\
071003 &  &  & $-0.97 \pm 0.07$ & $799^{+124}_{-100}$ & \markcite{GCN6849}{Golenetskii} {et~al.} (2007a) \\
071010B & $-1.25^{+0.74}_{-0.49}$ & $-2.65^{+0.29}_{-0.49}$ & & $52^{+14}_{-10}$ & \markcite{GCN6879}{Golenetskii} {et~al.} (2007b) \\
071020 & & & $-0.65^{+0.27}_{-0.32}$ & $322^{+80}_{-53}$ & \markcite{GCN6960}{Golenetskii} {et~al.} (2007c)\\
071117 & & & $-1.53^{+0.15}_{-0.16}$ & $278^{+236}_{-79}$ & \markcite{GCN7114}{Golenetskii} {et~al.} (2007e)\\
\enddata
\tablenotetext{*}{Short Hard GRB}
\tablecomments{GRBs with measured redshifts and $\gamma$-ray spectral
  properties from the literature used to calculated $E_{\gamma,iso}$.  CPL
  provided from either Band function fit or CPL fit, but $\alpha_{CPL}$ used
  only for SHBs.}
\end{deluxetable}
\renewcommand{\thefootnote}{\arabic{footnote}}

\end{document}